\documentclass[aps,pra,superscriptaddress,twocolumn,longbibliography]{revtex4-2}

\usepackage[english]{babel}
\usepackage{amssymb}
\usepackage{amsmath}
\usepackage{mathtools}
\usepackage{algorithm}
\usepackage{algpseudocode}
\usepackage{graphicx}
\usepackage{color}
\usepackage[colorlinks=true,
			urlcolor=blue,
			citecolor=blue,
			linkcolor=blue,
			citebordercolor={1 0 0},
			linkbordercolor={0 0 1}]{hyperref}
\usepackage{quantikz}
\usepackage{amsthm}

\usepackage{nameref}
\newcommand{\figref}[1]{Fig.~\ref{#1}}

\algrenewcommand\algorithmicrequire{\textbf{Input:}}
\algrenewcommand\algorithmicensure{\textbf{Output:}}

\newcommand{\bea}{\begin{eqnarray}}
\newcommand{\eea}{\end{eqnarray}}
\newcommand{\be}{\begin{equation}}
\newcommand{\ee}{\end{equation}}
\newcommand{\ba}{\begin{equation}\begin{aligned}}
\newcommand{\ea}{\end{aligned}\end{equation}}

\newtheorem{remark}{Remark}

\def\1{\mathds{1}}

\newcommand{\ketbra}[2]{\vert #1 \rangle \langle #2 \vert}

\makeatletter
\renewcommand\frontmatter@abstractwidth{\dimexpr\textwidth -0in \relax}

\usepackage{circuitikz}
\usepackage[normalem]{ulem} % normalem: keeps \emph italic (book titles in the reference list)

\usepackage{tikz,xcolor,hyperref}
\definecolor{lime}{HTML}{A6CE39}
\DeclareRobustCommand{\orcidicon}{%
	\begin{tikzpicture}
	\draw[lime, fill=lime] (0,0)
	circle [radius=0.16]
	node[white] {{\fontfamily{qag}\selectfont \tiny ID}};
	\draw[white, fill=white] (-0.0625,0.095)
	circle [radius=0.007];
	\end{tikzpicture}
	\hspace{-2mm}
}

\foreach \x in {A, ..., Z}{%
	\expandafter\xdef\csname orcid\x\endcsname{\noexpand\href{https://orcid.org/\csname orcidauthor\x\endcsname}{\noexpand\orcidicon}}
}

\newif\ifrevmark \revmarktrue
\newcommand{\rev}[1]{\ifrevmark{\color{black}#1}\else#1\fi}
\colorlet{BLUE}{blue} 

\begin{document}

\title{Rate-independent Defect Plateau in Quantum Many-Body Systems}

\author{Thi Ha Kyaw \orcidA{}}
\email{thiha.kyaw@lge.com}
\affiliation{LG Electronics Toronto AI Lab, Toronto, Ontario M5V 1M3, Canada}

\author{Guillermo Romero \orcidB{}}
\email{guillermo.romero@usach.cl}
\affiliation{Departamento de F\'isica, CEDENNA, Universidad de Santiago de Chile (USACH), Avenida Víctor Jara 3493, 9170124 Santiago, Chile}

\author{Gaurav Saxena \orcidC{}}
\affiliation{LG Electronics Toronto AI Lab, Toronto, Ontario M5V 1M3, Canada}

\date{\today}

\begin{abstract}
Adiabatic quantum computing is limited by the energy gap: the evolution time has to grow as the inverse square of the minimum gap, or the adiabatic approximation fails.
Shortcuts to adiabaticity speed up the evolution, but they need an additional Hamiltonian that is hard to implement.
Here we study a route without any extra Hamiltonian, in which the control follows a geodesic of the ground-state manifold and the Hamiltonian is applied as a train of global pulses.
Specifically, we consider many-body quantum systems and find that jumping along geodesics (geo-jump) suppresses the density of excitations at fast quenches.
We demonstrate analytically and numerically a rate-independent defect plateau: the quench rate drops out of the defect density, and the Kibble--Zurek dependence is carried by the number of pulses instead.
The plateau value is given by an exact mode-product expression; it lies below the sudden-quench value and, beyond a few kicks, decreases with their number, so that the quench time is traded for the number of pulses.
The geo-jump strategy is thus a hardware-frugal route to
state preparation that delivers the same local physical state over arbitrary runtimes, using only the native
Hamiltonian, with no auxiliary counterdiabatic fields.
\end{abstract}

\maketitle

\section{Introduction}

Adiabatic quantum computing (AQC) \cite{Farhi2000Jan,Albash2018Jan} is one of the standard models of quantum computation, and it is universal \cite{Aharonov2007}: it can perform any computation that the gate-based model can.
A problem is encoded in the ground state of a Hamiltonian, and the system is driven from an easily prepared initial ground state to the final one, which holds the solution, provided the evolution stays adiabatic.

AQC has been applied to combinatorial optimization problems in logistics \cite{ArinoSales2023}, finance \cite{Herman2022Jan}, machine learning \cite{Ma2023}, and materials science \cite{Camino2023Jun}, where the mapping of a cost function onto a Hamiltonian is direct and tunneling can help the search through the solution space.
Quantum annealers and quantum simulators now run such algorithms routinely.

The limitation of AQC is the energy gap. By the adiabatic theorem the evolution has to be slow enough to avoid transitions out of the instantaneous ground state, and the usual estimate of the required time grows as the inverse square of the minimum gap between the ground and first excited states \cite{Albash2018Jan}. When that gap closes exponentially with the system size the method becomes impractical.
The gap condition, however, is neither sufficient nor necessary for adiabaticity \cite{Marzlin2004Oct,Tong2005Sep}.

Several routes around the gap constraint have been explored. Shortcuts to adiabaticity (STA) \cite{Guery-Odelin2019Oct,Zhang2015Dec,Kyaw2018Apr}, such as counterdiabatic or transitionless driving, let a system follow the adiabatic path in a much shorter time by adding a Hamiltonian term that cancels the diabatic transitions.
The price is that term. Generating and controlling it in an experiment is hard, although not impossible~\cite{Bason2012,Romero2025,Visuri2026Digitized}; digitized versions of such terms have been run on IBM Q devices~\cite{Romero2025,Visuri2026Digitized}. For many-body systems the term is in general nonlocal, which limits how far STA methods scale.

Methods that suppress transitions between eigenstates without a long evolution time are therefore of interest \cite{Grabarits2025,Juhasz2023}, for adiabatic control in quantum computing and in atomic, molecular, optical, and condensed-matter physics.
Refs.~\cite{Wang2016May, Xu2019Jun} showed that adiabaticity is governed by the integral of the dynamical phase difference between the ground and first excited states, and not by a comparison between the evolution time and the gap (see Sec.~\ref{sec:gac} and Appendix~\ref{supp:adiabatic_condition}). Based on this condition, these works proposed to accelerate adiabatic evolution by applying sequences of $\pi$ pulses while traversing geodesics of the parameter space (``jumping along geodesics,'' hereafter geo-jump). The method was demonstrated experimentally in Ref.~\cite{Xu2019Jun} and has recently been extended to three-level systems~\cite{Xing2026}.

\begin{figure}[t]
        \centering
        \includegraphics[width=1\linewidth]{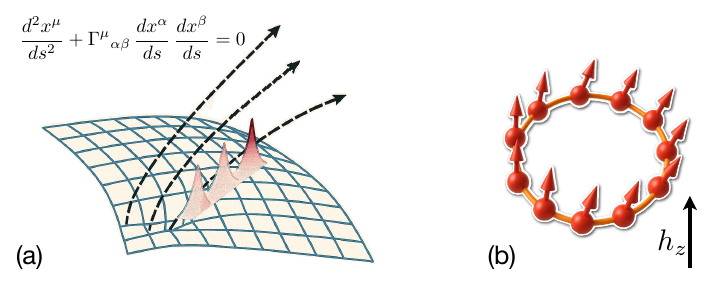}
        \caption{(a) An artistic impression of a quantum geodesic trajectory. The dashed arrows indicate the shortest paths, or geodesics, in a complex projective space endowed with a Hermitian form. The parameters $x$'s are associated with \rev{the} quantum system's external control parameters, and $s$ is associated with the evolution time $t$. $\Gamma$'s are the Christoffel symbols that are functions of \rev{the} Fubini--Study metric: $g_{\alpha \beta}$.
        A train of pulses indicates the means proposed in Refs.~\cite{Wang2016May,Xu2019Jun} to traverse the geodesic while minimizing quantum excitations in a relatively short evolution time, without needing to fulfill the standard adiabatic theorem.
        (b) \rev{The} XY spin model with $N$ quantum spins with periodic boundary conditions and external field $h_z$ \rev{considered in this study; it exhibits second-order phase transitions at various critical points}.
    }
        \label{fig:main_fig}
\end{figure}

In this paper we apply this method to many-body quantum systems, a case that, to our knowledge, has not been addressed before. This many-body extension, rather than the pulse-sequence idea itself, is the main contribution of the present work.
The method needs no auxiliary Hamiltonian. The control parameters follow a geodesic of the quantum state manifold \cite{Carinena2015, Zhang2023Jul}, as illustrated in Fig.~\ref{fig:main_fig}(a), and the evolution uses the original Hamiltonian only; the time-dependent controls come from the geodesic equation in the projective Hilbert space of the ground states, which for the one-parameter paths used here reduces to traversing the path at constant Fubini--Study speed (Sec.~\ref{sec:xy}).

\section{Generalized adiabatic condition}
\label{sec:gac}

Consider a smooth family of Hamiltonians $H(\lambda)$ with instantaneous eigenstates $\ket{\varphi_n(\lambda)}$ and energies $E_n(\lambda)$, where $\lambda=t/T\in[0,1]$ is the rescaled time and $T$ the total evolution time. A comparison of the true evolution with the ideal adiabatic transport of the eigenvectors gives the first-order off-diagonal amplitudes between instantaneous eigenstates $k\neq j$,
\begin{equation}
F_{kj}(\lambda)=\int_0^\lambda e^{iT[\alpha_k(\lambda')-\alpha_j(\lambda')]}\,G_{kj}(\lambda')\,d\lambda',
\label{eq:Fkj_main}
\end{equation}
where $\alpha_n(\lambda)=\int_0^\lambda E_n(\lambda')\,d\lambda'$ are the dynamical phases and $G_{kj}(\lambda)=i\,e^{-i[\gamma_k(\lambda)-\gamma_j(\lambda)]}\braket{\dot\varphi_k(\lambda)}{\varphi_j(\lambda)}$ is a slowly varying nonadiabatic coupling that absorbs the geometric phases $\gamma_n(\lambda)$. An integration by parts bounds $F_{kj}$ by the oscillatory integrals of the dynamical phase differences alone,
\begin{equation}
\epsilon_{kj}(\lambda):=
\left|
\int_0^\lambda e^{iT[\alpha_k(\lambda')-\alpha_j(\lambda')]}\,d\lambda'
\right|,
\label{eq:error}
\end{equation}
which we refer to as the error. Eigenstate transitions are suppressed to first order in the Dyson expansion provided
\begin{equation}
\sup_{\lambda\in[0,1]}\epsilon_{kj}(\lambda)\approx 0,
\quad \forall k\neq j.
\label{eq:gac_main}
\end{equation}
The derivation of Eqs.~\eqref{eq:Fkj_main}--\eqref{eq:gac_main} is given in Appendix~\ref{supp:adiabatic_condition}. The condition does not compare the evolution time with the energy gap. It can be met at short evolution times, and even when the gap closes, provided the dynamical phase factors add destructively. For a two-level system a train of $\pi$ pulses applied along a geodesic \rev{makes the dynamical phase factors of successive intervals cancel in Eq.~\eqref{eq:error}} and suppresses diabatic transitions however short the evolution \rev{is} (Sec.~\ref{sec:LZ}). In a many-body system the kick angle depends on the mode and the cancellation is partial; the resulting defect density is worked out in Appendix~\ref{supp:driven_bogoliubov}.

\section{Geo-jump strategy in a single qubit: the Landau--Zener model}
\label{sec:LZ}

Before turning to many-body systems, we illustrate the geo-jump strategy of Refs.~\cite{Wang2016May,Xu2019Jun} with the simplest example, the Landau--Zener (LZ) model.
At every point in time, the state of the two-level system (spin-1/2) can be modified by an external magnetic field. The LZ Hamiltonian reads
\begin{equation}
H_{LZ}(t) = \rev{x(t)X + \varepsilon Z},
\label{eq:HLZ}
\end{equation}
where $X$, $Z$ are the usual Pauli matrices. For our numerical experiments, we set $\varepsilon=0.1$ and the end points to $x_i=-10$ and $x_f=10$. The initial (final) state is the ground state of $H_{LZ}(x_i)$ [$H_{LZ}(x_f)$]. \rev{The instantaneous gap is $\Delta E_l=2\sqrt{x^2+\varepsilon^2}$, with its minimum $2\varepsilon=0.2$ at $x=0$.}

Finding a geodesic is in general not a simple task; Appendix~\ref{supp:geodesics} gives a short discussion. For a single qubit, great circles of the Bloch sphere are geodesics.
The starting point is an analytical expression for the ground-state wavefunction, parametrized here by the external controls $x$ and $\varepsilon$ of the qubit; see Appendix~\ref{supp:XYmodel} and Refs.~\cite{Cafaro2022Oct,Tomka2016Jun}.

\begin{figure*}[t]
        \centering
        \includegraphics[width=1.0\linewidth]{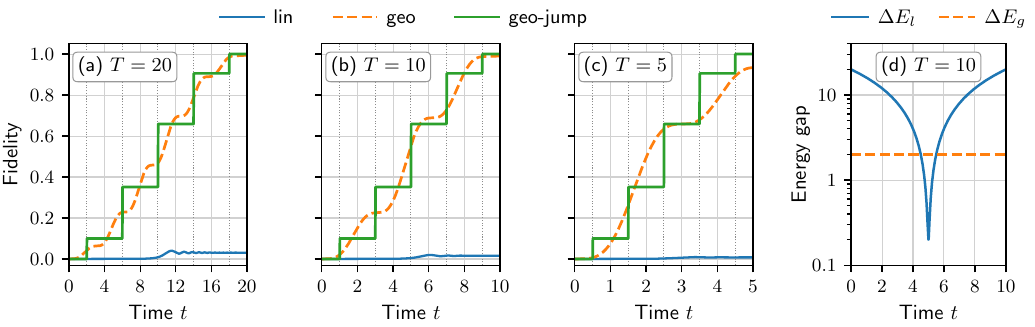}
        \caption{Unitary evolution under the LZ Hamiltonian, Eq.~\eqref{eq:HLZ}, for the lin, geo, and geo-jump strategies. The initial state is the ground state of $H_{LZ}(x_i)$ and the target state is the ground state of $H_{LZ}(x_f)$. The fidelity is the squared overlap between the target state and the instantaneous state evolved under each strategy. Panels (a)--(c) correspond to \rev{the total evolution times $T=20$, $10$, and $5$}. (d) shows\rev{, on a logarithmic scale,} the instantaneous energy gap between the ground and first excited states for the lin strategy ($\Delta E_l$) and the geo strategy ($\Delta E_g$), for the total evolution time of (b). The dependence on $\Delta E$ is visible in all three panels (a)--(c). For the longer evolution in (a) the system spends more time near the minimum of $\Delta E_l$, \rev{and the lin fidelity is larger; it has a small bump just after the anticrossing, around $t\approx11.5$, which shrinks in (b) and (c)}. The geo strategy works well in (a) and (b) because $\Delta E_g$ \rev{stays ten times above the minimum of} $\Delta E_l$; for the shorter evolution time of (c) it no longer reaches near-unit fidelity. The geo-jump strategy works in all three cases. \rev{It applies the same $R=5$ pulses in every panel, at the times $t_j=\lambda_jT$ marked by the dotted vertical lines. The fidelity (green) jumps where the $\pi$ pulses are applied and rises in the same five steps, to $0.100$, $0.351$, $0.658$, $0.906$, and $1$; these are the values that an adiabatic geo evolution would have at $t=jT/R$.}}
        \label{fig:LZ}
\end{figure*}

\begin{figure}[t]
        \centering
        \includegraphics[width=0.95\linewidth]{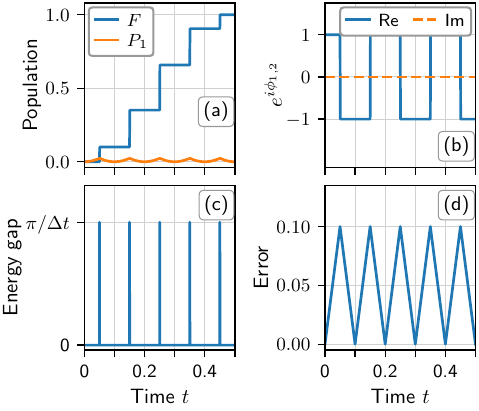}
        \caption{\rev{Geo-jump evolution of the LZ system for $T=0.5$ and $R=5$, with each pulse applied as a single time step of width $\Delta t=10^{-4}$; see the text for details. (a) Fidelity with the target state, $F=|\langle\psi(t_f)|\psi(t)\rangle|^2$, and population of the instantaneous excited state, $P_1=|\langle E_1(t)|\psi(t)\rangle|^2$. (b) Real and imaginary parts of $e^{i\phi_{1,2}}$, where $\phi_{1,2}$ is the dynamical phase difference. (c) Instantaneous energy gap of the pulsed Hamiltonian, which is zero between the pulses and $\pi/\Delta t$ during a pulse. (d) The error of Eq.~\eqref{eq:error}.} $|\psi(t_f)\rangle$ is the ground state of the final Hamiltonian, $|\psi(t)\rangle$ the instantaneous evolved state, and $|E_1(t)\rangle$ the instantaneous excited state \rev{of $H_{LZ,\mathrm{geo}}(t)$}.}
        \label{fig:LZphase}
\end{figure}

Throughout this work we compare three dynamical strategies. The linear (lin) strategy ramps the control parameter linearly in time; in the geodesic (geo) strategy the control parameter follows a predefined geodesic path; and the geo-jump strategy adds $\pi$ pulses to the geodesic trajectory. For the geo and geo-jump strategies, we define $\theta_{i/f}=\arctan(x_{i/f}/\varepsilon)$ and $\theta(t)=\theta_i+(\theta_f-\theta_i)t/T$. \rev{The two strategies use the LZ Hamiltonian normalized to unit field strength, $H_{LZ,\mathrm{geo}}=H_{LZ}/\sqrt{x^2+\varepsilon^2}=\sin\theta(t)X+\cos\theta(t)Z$ with $x=\varepsilon\tan\theta$, whose gap is $\Delta E_g=2$ at all times}. The geo-jump unitary evolution is defined by $H_{LZ,\mathrm{geo.jump}}=J_\pi(t)H_{LZ,\mathrm{geo}}$. $J_\pi(t)$ denotes a train of $\pi$ pulses equally spaced over the evolution, $J_\pi(t)=\frac{\pi}{2}\sum_{j=1}^{R}\delta(t-t_j)$, realized numerically as rectangular pulses of height $\pi/(2\Delta t)$ and width $\Delta t$, where $t_j=\lambda_j T$, $\lambda_j=(2j-1)/(2R)$, and $R$ is the number of pulses\rev{; we use $R=5$ throughout this section. Each pulse is the unitary $\exp[-i\frac{\pi}{2}H_{LZ,\mathrm{geo}}(t_j)]$, a rotation of the Bloch vector by $\pi$ about the field direction at $t_j$. It takes the ground state at $\lambda=(j-1)/R$ to the ground state at $\lambda=j/R$, so after the $j$th pulse the fidelity with the target state is $\cos^2[(1-j/R)(\theta_f-\theta_i)/2]$, whatever the value of $T$. For $R=5$ these values are $0.100$, $0.351$, $0.658$, $0.906$, and $1$}. The resulting fidelities for the lin, geo, and geo-jump strategies are shown in Fig.~\ref{fig:LZ} \rev{for $T=20$, $10$, and $5$. The lin ramp ends at the fidelities $0.031$, $0.016$, and $0.008$, which are the Landau--Zener values, and the geo ramp at $0.995$, $0.990$, and $0.934$, while the geo-jump staircase is the same in the three panels and ends at $1$}. The geo-jump strategy reaches the best fidelity independently of the total evolution time; it matches the geo strategy for long times and outperforms it for short ones.

To supplement the generalized adiabatic condition, Eq.~\eqref{eq:gac_main}, and the numerical results of Fig.~\ref{fig:LZ}, we include Fig.~\ref{fig:LZphase}\rev{, for the short evolution time $T=0.5$}. The energy-gap panel shows that $J_\pi(t)$ envelops the entire LZ Hamiltonian and that the gap is closed \rev{between the pulses, that is,} for most of the evolution. Since traversing from the initial state to the final target one gives \rev{a} constant geometric phase, the dynamical phase differences $\phi_{1,2}$ between the two instantaneous eigenstates become important. \rev{Each pulse advances $\phi_{1,2}$ by $\pi$, so $e^{i\phi_{1,2}}$ is real and changes sign at every pulse [Fig.~\ref{fig:LZphase}(b)]. The error of Eq.~\eqref{eq:error} then never exceeds $1/(2R)=0.1$, and it returns to zero at $\lambda=j/R$ [Fig.~\ref{fig:LZphase}(d)].} As a result, one can make the unitary as close to the adiabatic one as possible even when the energy gap closes. This is apparent from the fidelity \rev{reaching} one even when the total time is $0.5$\rev{, while the population of the excited state of $H_{LZ,\mathrm{geo}}(t)$ stays below $0.025$ [Fig.~\ref{fig:LZphase}(a)]}.

\section{Main theoretical result}
\label{sec:main}

\textit{Rate independence.} Let $H_\Lambda(s)$, $s\in[0,1]$, be a smooth path of local Hamiltonians
on a finite lattice region $\Lambda$, with uniformly bounded,
finite-range interaction terms and a uniform spectral gap
$\Delta>0$ above the (possibly degenerate) ground space. Consider the
geo-jump protocol consisting of $R$ $\delta$-kicks of fixed area
$a$ applied at fixed scaled times $s_j\in[0,1]$, generating
\begin{equation}
  U_{\Lambda,R} \;=\; \prod_{j=1}^{R} e^{-i a H_\Lambda(s_j)},
  \qquad \hbar = 1 .
\end{equation}
By construction, for any initial state
$\ket{\psi_\Lambda}$ and any bounded observable
$\mathcal{O}_\Lambda$, the post-protocol expectation value
\begin{equation}
  \langle \mathcal{O}_\Lambda \rangle_T
  \;=\;
  \bra{\psi_\Lambda} U_{\Lambda,R}^{\dagger}\,
  \mathcal{O}_\Lambda\, U_{\Lambda,R} \ket{\psi_\Lambda}
\end{equation}
is independent of the total evolution time $T$, since the kick times
$t_j = s_j T$ scale with $T$ while the kick areas do not. This rate independence is a direct consequence of the protocol, whose evolution operator is an ordered product of fixed unitaries. Between the kicks the Hamiltonian vanishes and the state does not evolve, so $T$ only sets the spacing of the kicks; the physical duration of the protocol is the total pulse width $R\,\Delta t$.

\textit{Plateau value (transverse-field Ising chain).} For the
Ising line of the XY model, Eq.~\eqref{eq:HXY}, with the transverse field $h$ ramped from $h_i$ to $h_f$ at anisotropy $\gamma$ and kicked $R$ times, the defect density is given by the exact mode-product expression, Eq.~\eqref{eq:exact_nk}. Each kick multiplies the ground and excited components of the Bogoliubov mode $k$ by $e^{\pm i\alpha_{kj}}$, with $\alpha_{kj}=\pi\varepsilon_k(h_j)$ set by the mode energy $\varepsilon_k=\sqrt{a_k^2+\Delta_k^2}$ at the kick field $h_j$ [Eq.~\eqref{eq:Hk_main}; $\varepsilon_k$ is written $E_k$ in Appendix~\ref{supp:driven_bogoliubov}]; that is, it rotates the mode by $2\alpha_{kj}$ about its instantaneous field. Writing every kick in its own eigenbasis gives two general properties (Appendix~\ref{supp:driven_bogoliubov}). First, the kicks never create more defects than a sudden quench (a consistency check),
\begin{equation}
n^{\mathrm{geo.jump}}_{\mathrm{defect}}\le n_{\mathrm{sudden}}=\frac{1}{\pi}\int_0^{\pi}dk\,\sin^2\frac{\Theta_k}{2},
\label{eq:bound_main}
\end{equation}
where $\Theta_k$ is the angle between the initial and final fields of mode $k$, for any $R$, any kick times, and any monotonic ramp. Second, for a small ramp, $\Theta_k\ll1$, the plateau decreases as $n^{\mathrm{geo.jump}}_{\mathrm{defect}}=C/R+O(R^{-2})$, with $C$ fixed by the modes whose kick angle is a multiple of $\pi$, Eq.~\eqref{eq:CR}.

The $T$-independence above is exact and structural;
the physical content of the plateau lies in its value, which never exceeds the sudden-quench value and, once $R$ exceeds a few, falls off with $R$, while the quench rate drops out altogether.
We further note that the locality and gap assumptions are not required
for the rate independence, which is purely algebraic and holds for arbitrary Hamiltonian
paths, initial states, and bounded observables; they delimit the
regime in which the plateau value is small and locally
determined.
The rate independence is shown in Appendix~\ref{supp:theorem1_proof}, and the plateau value is derived in Appendix~\ref{supp:driven_bogoliubov}.
The result shows that, for local gapped many-body systems, geo-jump dynamics eliminate the conventional dependence of local defects on evolution time. Consequently, the same local physical state can be prepared over a broad range of runtimes, with the number of pulses in place of the quench time, using only the native Hamiltonian and without auxiliary counterdiabatic fields.

\section{Geo-jump strategy in the XY model and the toric code}
\label{sec:xy}

\begin{figure*}[t]
        \centering        \includegraphics[width=1.0\linewidth]{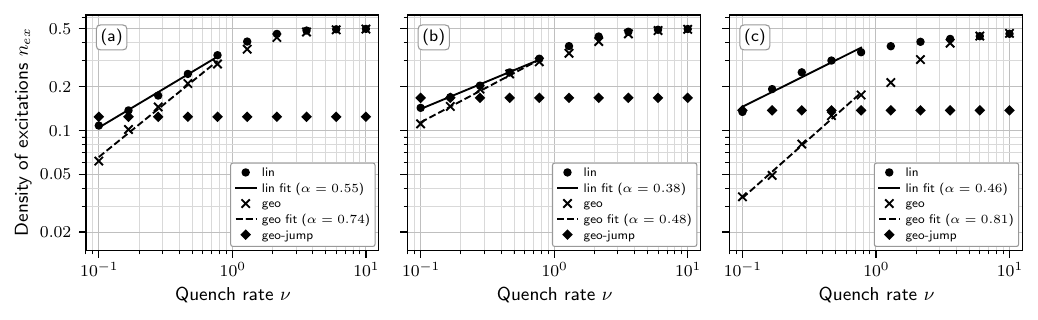}
        \caption{Density of excitations/defects $n_{ex}$ generated \rev{in the one-dimensional XY model with $N=250$ spins, plotted against the quench rate $\nu=1/T$ in a log-log plot}, for the lin (circles), geo (crosses), and geo-jump (diamonds, $R=5$) strategies.
        The \rev{XY} model has several critical lines, selected by the choice of parameters.
        (a) corresponds to the anisotropy line with $h=0.5$ with varying $\gamma(t)$ from $\gamma_i = -1$ \rev{to} $\gamma_f =1$.
        (b) is the gapless line with $h/J =1$ with varying $\gamma(t)$ from $\gamma_i = -1$ \rev{to} $\gamma_f =1$.
        (c) is the Ising line with fixed $\gamma =1$ and varying $h(t)$ from $h_i = 10$ to $h_f =0.$
        The geo and geo-jump strategies use the schedule shared by all modes, Eq.~\eqref{eq:closed_schedules}, and $n_{ex}$ is the average of \rev{the excitation probability} $p_k$ over the momenta $k=(2m-1)\pi/N$ of the chain. Solid (dashed) lines are least-squares fits of $\log n_{ex}$ against $\log\nu$ for the lin (geo) data over the slow-quench window $0.1\le\nu\le0.8$; the legends give the fitted effective exponents $\alpha$. The three panels share the vertical axis.
        From the figures, it is apparent that for the geo-jump strategy the density of excitations does not depend on the quench rate $\nu$ in all three cases.
    }
        \label{fig:KZ_scaling_results}
\end{figure*}

Implementing the protocol requires the geodesics of the many-body system \cite{Tomka2016Jun}.
To illustrate the method, we consider two quantum models: the XY quantum spin model \rev{in a transverse field~\cite{Lieb1961,Katsura1962,Barouch1970,Barouch1971}} and the toric code\rev{~\cite{Kitaev2003,Dennis2002}}.
We first investigate the XY model [Fig.~\ref{fig:main_fig}(b)] numerically with $N=250$ spins across three different parameter regimes (see \figref{fig:KZ_scaling_results}).
The Hamiltonian is given by
\begin{multline}
    H_{XY} = -J\sum_{j=1}^N \bigg[\frac{(1+\gamma)}{2} X_j X_{j+1}\\
    + \frac{(1-\gamma)}{2}Y_j Y_{j+1}+ hZ_j \bigg],
\label{eq:HXY}
\end{multline}
where $X,Y,$ and $Z$ are the standard Pauli operators, and $\gamma$ and $h$ are external control parameters; we set $J=1$.
\rev{The model is exactly solvable: a Jordan--Wigner transformation maps it onto free fermions~\cite{Lieb1961,Katsura1962,Pfeuty1970,Barouch1970}, and its phase diagram and quench dynamics are reviewed in Ref.~\cite{Dutta2015}. The} geodesic trajectories can \rev{therefore} be determined explicitly in terms of the control parameters $J(t), \gamma(t), h(t)$.
Depending on the choice of the parameters, three different scenarios can be considered.
The anisotropy line ($|h|<|J|$) and the gapless line ($h/J=1$) can be obtained with time-dependent $\gamma(t)$, whereas by choosing $\gamma=1$ with time-varying $h(t)$, the model converges to the Ising model.
For the three cases mentioned, the corresponding geodesic connecting the chosen initial and final many-body wavefunction can be constructed as follows; the detailed calculation is given in Appendix~\ref{supp:XYmodel}. After the Jordan--Wigner transformation followed by the momentum-space representation, the XY Hamiltonian decouples into independent modes $k$, each described by the $2\times 2$ Hamiltonian
\begin{equation}
\begin{gathered}
H_k=-2\left(a_k\,\sigma_z+\Delta_k\,\sigma_x\right),\\
a_k=h-\cos k,\qquad \Delta_k=\gamma\sin k .
\end{gathered}
\label{eq:Hk_main}
\end{equation}
Its ground state, $\ket{\psi_0}=\cos(\theta_k/2)\ket{\uparrow_k}+\sin(\theta_k/2)\ket{\downarrow_k}$, is fixed by the mixing angle $\tan\theta_k=\Delta_k/a_k$, so that the Fubini--Study metric of each mode reduces to $g^k_{\lambda\lambda}=\tfrac{1}{4}\left(\partial\theta_k/\partial\lambda\right)^2$, with $\lambda$ the time-dependent control parameter ($\gamma$ or $h$). The geodesic condition $g^k_{\lambda\lambda}\dot\lambda^2=\mathrm{const.}$~\cite{Tomka2016Jun} then implies that the mixing angle advances at a constant rate,
\begin{equation}
\theta_k(t)=\theta_{k,i}+\left(\theta_{k,f}-\theta_{k,i}\right)\frac{t}{T},
\label{eq:theta_main}
\end{equation}
with $\theta_{k,i}$ and $\theta_{k,f}$ the mixing angles of the initial and final Hamiltonians. Inverting $\tan\theta_k=\Delta_k/a_k$ yields the geodesic control schedules,
\begin{equation}
\begin{aligned}
\gamma(t)&=\frac{h-\cos k}{\sin k}\,\tan\theta_k(t),\\
h(t)&=\cos k+\gamma\sin k\,\cot\theta_k(t),
\end{aligned}
\label{eq:geo_schedules}
\end{equation}
for the anisotropy and gapless lines, and for the Ising line, respectively.

Equation~\eqref{eq:geo_schedules} gives a different schedule for every mode $k$, because the mixing angles depend on $k$. Such a mode-resolved schedule can be realized in the mode-by-mode emulation discussed in Sec.~\ref{sec:outlook}, but a spin chain has a single control field. For the many-body protocol we therefore use the ground state of the whole chain and traverse the one-parameter curve of its ground states at constant Fubini--Study speed, which is the geodesic protocol of Appendix~\ref{supp:Fubini--Study} applied to the chain as a whole. The ground state is a product over the modes $k=(2m-1)\pi/N$, $m=1,\dots,N/2$, so its metric along the control line is the sum of the mode metrics,
\begin{equation}
g_{\lambda\lambda}=\sum_{k>0}g^{k}_{\lambda\lambda}=\frac{1}{4}\sum_{k>0}\left(\frac{\partial\theta_k}{\partial\lambda}\right)^{2},
\label{eq:global_metric}
\end{equation}
and the geodesic protocol $g_{\lambda\lambda}\dot\lambda^2=\mathrm{const.}$ fixes one schedule $\lambda(t)$ for all modes, the inverse of the normalized arc length, Eq.~\eqref{eq:global_schedule}. On a one-parameter line all monotonic paths have the same length, so this condition fixes the parametrization only; the control moves at constant speed in the fidelity-susceptibility metric, a protocol related to the quantum adiabatic brachistochrone of Ref.~\cite{Rezakhani2009} (see also Ref.~\cite{Kolodrubetz2017}). In the thermodynamic limit the sum becomes an integral over $k$ and has a closed form on both lines (Appendix~\ref{supp:global_schedule}). The schedule is
\begin{equation}
\begin{aligned}
h(s)&=\begin{cases}1/\cos[(s_c-s)L], & s\le s_c,\\ \cos[(s-s_c)L], & s\ge s_c,\end{cases}\\
\gamma(s)&=\mathrm{sgn}(a)\,\tan^{2}a,\qquad a=\tfrac{\pi}{2}\left(s-\tfrac{1}{2}\right),
\end{aligned}
\label{eq:closed_schedules}
\end{equation}
with $s=t/T$, on the Ising line ($h\colon10\to0$, with $L=\arccos(1/10)+\pi/2$ and $s_c=\arccos(1/10)/L=0.484$) and on the anisotropy and gapless lines ($\gamma\colon-1\to1$, where the metric does not depend on $h$ for $|h|\le1$), respectively. Near the critical point the metric diverges as $1/|\lambda-\lambda_c|$ and the schedule slows down: it crosses the critical point as $|\lambda-\lambda_c|\propto(t-t_c)^2$, a quadratic crossing, whereas the lin ramp keeps a constant speed. All the results below use this shared schedule for the geo and geo-jump strategies; the mode-resolved schedule of Eq.~\eqref{eq:geo_schedules} is compared in Appendix~\ref{supp:additional_numerics}.

For each case, we implement the three dynamical strategies introduced in Sec.~\ref{sec:LZ}: lin, geo, and geo-jump, with $R=5$ kicks of area $\pi/2$ at the scaled times $\lambda_j=(2j-1)/(2R)$ in the geo-jump strategy. Each kick is $\exp[-i\frac{\pi}{2}H_{XY}(\lambda_j)]$, a ``$\pi$ pulse'' in the sense of Refs.~\cite{Wang2016May,Xu2019Jun}: its area is fixed, and it rotates mode $k$ by the mode-dependent angle $2\pi\varepsilon_k(\lambda_j)$, where $\varepsilon_k=\sqrt{a_k^2+\Delta_k^2}$, i.e., $\alpha_{kj}=\pi\varepsilon_k(\lambda_j)$ in the notation of Appendix~\ref{supp:driven_bogoliubov}.
In Fig.~\ref{fig:KZ_scaling_results} the lin strategy follows the Kibble--Zurek (KZ) power law $n_{ex}\propto\nu^{\alpha}$ \rev{in the quench rate $\nu=1/T$} at slow quenches, with effective exponents $\alpha=0.55$, $0.38$, and $0.46$ fitted over $0.1\le\nu\le0.8$ in panels (a), (b), and (c). The asymptotic KZ exponents of these three paths are $1/2$~\cite{Mukherjee2007}, $1/3$~\cite{Divakaran2008}, and $1/2$~\cite{Dziarmaga2005}. The window $0.1\le\nu\le0.8$ is still a crossover regime: when the same calculation is pushed to $\nu=10^{-3}$ in the thermodynamic limit, the local exponents settle at $0.50$, $0.33$, and $0.50$ (Appendix~\ref{supp:KZM} and Fig.~\ref{fig:slow}).
With the geo strategy the defect densities are smaller and the effective exponents larger, $\alpha=0.74$, $0.48$, and $0.81$. The shared schedule crosses the critical point as $|\lambda-\lambda_c|\propto(t-t_c)^2$, and on the Ising and anisotropy lines the Kibble--Zurek exponent for such a nonlinear crossing, $2/3$~\cite{Sen2008,Barankov2008}, is reached at slower rates (Table~\ref{table:summary}, Appendix~\ref{supp:KZM}, and Fig.~\ref{fig:slow}).
The geo-jump strategy, in contrast, gives a defect density that does not change with the quench rate, a rate-independent defect plateau (DRIP) at $n_{ex}=0.124$, $0.167$, and $0.137$ in the three cases. The plateau lies below the continuous ramps only at fast quenches: in Fig.~\ref{fig:KZ_scaling_results} it crosses the lin curves near $\nu\simeq0.1$--$0.2$ and the geo curves near $\nu\simeq0.2$--$0.5$. Since every kick has area $\pi/2$, the $R$ kicks carry the same integrated action as a continuous ramp of duration $\pi R/2$; for $R=5$ that geodesic ramp ($T\simeq7.9$) gives $n_{ex}\simeq0.04$ on the Ising line, against the plateau $0.137$. The protocol trades the duration of the ramp for the number of pulses, and it is the duration that drops out while the total pulse area stays fixed.
Defect density versus a much larger range of quench rate can be seen in \figref{fig:long_quench_rates_250spins}.
A plateau in defect density is also predicted in adiabatic transitions assisted by an auxiliary long-range Hamiltonian \cite{PhysRevLett.109.115703}.
This is achieved, for example, by adding an additional Hamiltonian which is the sum of $N/2$-body interacting terms for \rev{an} $N$-spin Ising chain.
In contrast, the geo-jump protocol multiplies the system Hamiltonian by a train of $\pi$ pulses, $H^{\textrm{geo.jump}}(t)=J_\pi(t)\,H^{\textrm{geo}}(t)$, with $J_\pi(t)$ the pulse train of Sec.~\ref{sec:LZ}: a rectangular pulse of height $\pi/(2\Delta t)$ in each interval $t\in[t_j,t_j+\Delta t]$, $t_j=\lambda_jT$.

The detailed derivation of the plateau value appears in Appendix~\ref{supp:driven_bogoliubov}. Each kick applies a fixed pulse area $\pi/2$ of the global Hamiltonian and rotates mode $k$ by $2\alpha_{kj}$, $\alpha_{kj}=\pi\varepsilon_k(h_j)$, about its instantaneous field. In the frame that follows this field, a kick only adds opposite phases to the ground and excited components; the transitions come from the rotation of the frame between kicks, weighted by the phases the kicks have accumulated. This is the kicked version of the generalized adiabatic condition of Sec.~\ref{sec:gac}; the sudden-quench bound, Eq.~\eqref{eq:bound_main}, and the decrease with $R$ stated in Sec.~\ref{sec:main} follow from it (Appendix~\ref{supp:driven_bogoliubov}). For the shared schedule across the critical point the critical modes follow the Kibble--Zurek law of the continuous ramp with $\pi R/2$ in place of $\tau_Q$, i.e., $R^{-2/3}$ for the quadratic crossing of Eq.~\eqref{eq:closed_schedules}, which the total approaches only at large $R$ (Appendix~\ref{supp:nR} and Fig.~\ref{fig:nR}).
The main results from the Ising model are summarized in Table~\ref{table:summary}.

\textit{Toric code.} \rev{The toric code~\cite{Kitaev2003} is a model of qubits on the edges of a square lattice on a torus. Its Hamiltonian, Eq.~\eqref{eq:TC-Hamiltonian}, is a sum of mutually commuting four-qubit star and plaquette operators with couplings $J_e$ and $J_m$. The ground space is fourfold degenerate, and the excitations are pointlike anyons, which are the defects of this model (Appendix~\ref{supp:toric_code}).} For the toric code driven along a commuting-projector path (varying only the couplings $J_e(s)$ and $J_m(s)$), the geo-jump protocol $U_R$ preserves the entire ground-space manifold $\mathrm{Ran}\,P(0)$ \rev{(the range of the ground-space projector at $s=0$)} and acts on it only by a \rev{common} phase. Hence, no anyon defects are created from any initial ground state or ground-space superposition. This holds for any protocol along such a path, a sudden quench included, and is a consistency check rather than a property of the kicks.
See Appendix~\ref{supp:toric_code}.

\begin{table*}[t]
\setlength{\tabcolsep}{0.45em}
\small
\begin{center}
\caption{Summary of the three strategies and their respective defect scaling results for the \rev{one-dimensional} Ising model, $h\colon10\to0$ at $\gamma=1$; $\tau_Q=T/|h_f-h_i|$ is the quench time in the convention of Ref.~\cite{Dziarmaga2005}. The exponents are the slow-quench (asymptotic) values; the effective exponents fitted over $0.1\le\nu\le0.8$ in Fig.~\ref{fig:KZ_scaling_results}(c) are $0.46$ and $0.81$. The $R$ dependence of the geo-jump plateau is shown in Fig.~\ref{fig:nR}.}
\label{table:summary}
\begin{tabular}{lcc}
\hline
\textbf{Strategy} & \textbf{Control} & \textbf{Defect Scaling} \\
\hline
Lin (KZ) & $h(t) = h_i + (h_f - h_i)\,t/T$
& $n_{\mathrm{defect}} ^{\rm KZ} \propto \tau_Q^{-1/2}$ \\[6pt]
Geo & $h(s)$ of Eq.~\eqref{eq:closed_schedules} (constant Fubini--Study speed)
& $n_{\mathrm{defect}}^{\rm geo} \propto \tau_Q^{-2/3}$ (quadratic crossing, $r=2$)
\\[6pt]
Geo-jump & $R$ kicks of pulse area $\pi/2$ at $\lambda_j = \tfrac{2j-1}{2R}$
& $n_{\mathrm{defect}} ^{\rm geo.jump}$ independent of $\tau_Q$; $0.137$ at $R=5$, decreasing with $R$ \\[4pt]
\hline
\end{tabular}
\end{center}
\end{table*}
\section{Experimental outlook}
\label{sec:outlook}

We see two routes to realizing the geo-jump strategy in many-body quantum systems. The first
exploits the free-fermion structure of the models considered here. After the Jordan--Wigner and Fourier transformations, the dynamics
decouples into independent Bogoliubov $k$ modes, each evolving as an effective two-level system in a time-dependent field. One may therefore
emulate each $k$ mode with a single qubit, sweeping its control along the shared schedule of Eq.~\eqref{eq:global_schedule}, or along the mode-resolved geodesic of Eq.~\eqref{eq:geo_schedules}, which is available in this setting because every mode is a separate qubit (Appendix~\ref{supp:additional_numerics}), and applying the prescribed $\pi$ pulses; the many-body defect density is then
reconstructed by summing the measured excitation probabilities $p_k$ over the sampled momenta, in direct analogy to earlier mode-by-mode quantum simulations of Kibble--Zurek physics \cite{Xu2014,Cui2016,Gong2016,Cui2020}. Because each mode requires only single-qubit control and readout, this route is open on any current platform, superconducting, trapped-ion, photonic, or semiconductor, and it allows the predicted collapse of the defect density across quench rates to be tested mode by mode. The second route addresses the many-body problem directly. One practical feature
of the geo-jump protocol is that the $\pi$ pulses multiply the global system Hamiltonian: the kicks are applied collectively and homogeneously to all spins, and no site-resolved addressing is
required. Global control of this kind is native to trapped-ion crystals and to neutral-atom arrays, where uniform Rabi drives and global detuning ramps are routinely engineered with high fidelity, and where Kibble--Zurek physics has already been probed experimentally \cite{Bernien2017}. The protocol also requires the full interacting Hamiltonian to be switched off between the pulses and on, at amplitude $\pi/(2\Delta t)$, during them. In trapped-ion crystals the spin-spin coupling is laser-mediated and can be applied during the pulses only; in neutral-atom arrays the Rydberg interaction is on whenever the atoms are excited, so the pulsed protocol would need the interaction to be gated, as in Rydberg-dressing schemes. The requirements are (i) the ability to ramp the control parameter along the precomputed geodesic schedule and (ii) pulse widths $\Delta t$ short compared with the inverse of the relevant
many-body energy scales, so that the kicks remain in the $\delta$-function regime; the sensitivity of the plateau to finite pulse width is quantified in Appendix~\ref{supp:additional_numerics} (Fig.~\ref{fig:PkDt}), and sets the accuracy budget for such an experiment. Requirement (ii) is the price of dispensing with auxiliary Hamiltonians. The geo-jump strategy replaces additional, often multi-body, control terms by short global pulses, and a finite pulse width reintroduces a dependence of the excitation probabilities on the quench rate (Fig.~\ref{fig:PkDt}). In the $\delta$-kick limit the plateau holds over four decades of the quench rate (Fig.~\ref{fig:long_quench_rates_250spins}). Superconducting annealers \cite{Bando2020,King2022}, which offer fast global schedule control, provide a further natural testbed for the rate independence reported here.

\section{Summary and discussion}

The geo-jump strategy of Refs.~\cite{Wang2016May,Xu2019Jun} consists of jumping along a geodesic of the quantum state manifold while applying $\pi$ pulses. For a single qubit the pulses \rev{make the dynamical phase factors of successive intervals cancel}, so diabatic transitions are suppressed at any evolution time (Sec.~\ref{sec:LZ}). The relevant timescale is not set by the instantaneous energy gap. For the qubit, the gap may close several times along the geo-jump trajectory without loss of fidelity, because the interstate transitions are governed by the integrated dynamical phase difference (Appendix~\ref{supp:adiabatic_condition}). In the chain the kick angle varies from mode to mode, so the cancellation is only partial, and the plateau is the interference of the frame rotations between kicks, Eq.~\eqref{eq:first_order}, summed over the modes. On the analytical side, we have shown that the geo-jump protocol yields a rate-independent defect plateau across the critical point of the one-dimensional transverse-field Ising model (see Table~\ref{table:summary}), with the plateau value given by the exact mode-product expression, Eq.~\eqref{eq:exact_nk}. This value is bounded by the sudden-quench defect density, Eq.~\eqref{eq:bound_main}, and for more than a few kicks it decreases with $R$, as $C/R$ for small ramps, Eq.~\eqref{eq:CR}. For the shared schedule across the critical point the critical modes follow the Kibble--Zurek law of the continuous ramp with $\pi R/2$ in place of $\tau_Q$, $\propto R^{-2/3}$ (Appendix~\ref{supp:nR}). On the numerical side, we have confirmed the rate independence of the defect density in three regimes of the XY model (the anisotropy line, the gapless line, and the Ising line) for chains of $N=250$ spins, large enough that finite-size effects are negligible on the scales probed (the $R=5$ plateau on a fine momentum grid is $0.1371$, as on the chain momenta); in all three cases the linear ramps follow Kibble--Zurek power laws, with effective exponents within $0.05$ of the asymptotic values, while the geo-jump data collapse onto a quench-rate-independent plateau (Fig.~\ref{fig:long_quench_rates_250spins}). The quench $h_i=10\to h_f=0$ used in the numerics is not a small ramp, so these data test the rate independence of the plateau and its dependence on $R$, Fig.~\ref{fig:nR}, rather than the small-ramp law. For topologically ordered
systems, exemplified by the toric code, we have further shown that, along commuting-projector paths, the protocol preserves the entire ground-space manifold and creates no
anyonic defects, as any protocol along such a path does. Analytical treatments and numerical experiments for other one-dimensional models, including \rev{nonintegrable} ones, are left
for future studies, as is the generalization to two dimensions, where geodesics of the quantum geometric tensor are no longer available in
closed form and must be constructed variationally or numerically.

For adiabatic quantum computation, the geo-jump protocol removes the quench time from the defect density without any counterdiabatic term of the shortcuts-to-adiabaticity type \cite{Zhang2015Dec,Kyaw2018Apr,Romero2025,Visuri2026Digitized}, and it is not tied to the instantaneous gap along the path. Assisted-adiabatic schemes reach a defect plateau by adding long-range multi-body interactions \cite{PhysRevLett.109.115703}, a sum of $N/2$-body terms for an $N$-spin Ising chain; the present proposal needs only the native Hamiltonian and a train of global $\pi$ pulses. The result is not a way around gap-limited runtime bounds. The geo-jump schedule requires the geodesic of the ground-state manifold, which amounts to detailed spectral information about the model, available in closed form here because the models are integrable. The protocol trades evolution time for classical precomputation, control complexity, and the number of pulses; it does not remove the difficulty of the problem. Nonetheless, a defect density that does not depend on the quench rate, and whose critical-mode contribution follows the Kibble--Zurek law with the number of kicks in place of the quench time, is an experimentally testable prediction. It shows that a train of global kicks removes the quench time from the defect density without a counterdiabatic term; what remains is the dependence on the number of kicks. State preparation that does not depend on the runtime is of use in annealing and in analog simulation.\\

\begin{acknowledgments}
We thank Jiang Zhang, Tonghao Xing and Jack S. Baker for fruitful discussions. We extend our thanks to Kevin Ferreira, Yipeng Ji, Paria Nejat of LG Electronics Toronto AI Lab and Sean Kim of LG Electronics, Next-generation Computing Lab, for their constant administrative support. G.R. acknowledges the Center for Nanoscience and Nanotechnology, CEDENNA, Project CIA250002, Proyecto de Exploración Grant No. 13250014, and DICYT Regular 042631RH{\_}REG Vicerrectoría de Investigación, Innovación y Creación, USACH.
\end{acknowledgments}

\appendix

\section{Derivation of the generalized adiabatic condition \label{supp:adiabatic_condition}}

In this appendix we derive the generalized adiabatic condition discussed in Refs.~\cite{Wang2016May,Xu2019Jun}.
We consider a finite-dimensional, smooth family of Hamiltonians
\(
H(\lambda)\) with \(\lambda\in[0,1]\).
For each \(\lambda\) let
\begin{equation*}
\begin{gathered}
H(\lambda)\,\ket{\varphi_n(\lambda)}
  = E_n(\lambda)\,\ket{\varphi_n(\lambda)},
\\
\braket{\varphi_m(\lambda)}{\varphi_n(\lambda)}=\delta_{mn},
\end{gathered}
\end{equation*}
and assume that the spectrum is \rev{nondegenerate} and gapped:
\begin{equation*}
\begin{gathered}
\Delta_{mn}(\lambda)
 := E_m(\lambda)-E_n(\lambda),
\\
|\Delta_{mn}(\lambda)|\ge g>0
\quad\forall\,m\neq n,\ \lambda\in[0,1].
\end{gathered}
\end{equation*}
We also assume that \(H(\lambda)\) and the eigenvectors
\(\ket{\varphi_n(\lambda)}\) are \(C^2\) in \(\lambda\).
The physical time is \(t=T\lambda\) with total \rev{runtime}
\(T>0\).
The true Schr\"odinger evolution \(U_T(\lambda)\) in the
\(\lambda\)-parametrization satisfies
\begin{equation}
i\,\frac{d}{d\lambda}U_T(\lambda)
  = T\,H(\lambda)\,U_T(\lambda),
\qquad
U_T(0)=\mathbb{I}.
\label{eq:lambda_Schrodinger}
\end{equation}

Let us define an operator \(A(\lambda)\) that maps
the eigenbasis at \(\lambda=0\) to that at \(\lambda\), including both
dynamical and geometric phases:
\begin{equation*}
\begin{aligned}
\alpha_n(\lambda)
   &:= \int_0^\lambda E_n(\lambda')\,d\lambda',
\\
\gamma_n(\lambda)
   &:= i\int_0^\lambda
      \braket{\varphi_n(\lambda')}{\dot\varphi_n(\lambda')}\,d\lambda'
\end{aligned}
\end{equation*}
(where the dot denotes \(d/d\lambda\)), and the phase-dressed
eigenvectors
\begin{equation*}
\ket{\psi_n(\lambda;T)}
  := e^{-iT\alpha_n(\lambda)+i\gamma_n(\lambda)}
     \ket{\varphi_n(\lambda)}.
\end{equation*}
By construction the set \(\{\ket{\psi_n(\lambda;T)}\}_n\) is
orthonormal for each \(\lambda\).
We now define the unitary
\begin{equation}
A(\lambda)
  := \sum_n \ket{\psi_n(\lambda;T)}\bra{\varphi_n(0)}.
\end{equation}
It implements the ideal adiabatic transport of eigenvectors:
\(A(\lambda)\ket{\varphi_n(0)}=\ket{\psi_n(\lambda;T)}\).

We compare the true evolution with this ideal evolution by
\begin{equation}
W_T(\lambda) := A^\dagger(\lambda)\,U_T(\lambda).
\label{eq:WT}
\end{equation}
Clearly \(W_T(0)=\mathbb{I}\).
The adiabatic theorem, in this language, will follow from
\begin{equation}
\begin{gathered}
\lim_{T\to\infty} W_T(\lambda)=\mathbb{I}
\\
\text{(in operator norm, uniformly in }\lambda\in[0,1]\text{)},
\end{gathered}
\label{eq:what_we_want_to_prove}
\end{equation}
which is equivalent to
\(\lim_{T\to\infty}A^\dagger(\lambda)U_T(\lambda)=\mathbb{I}\)
and hence \(U_T(\lambda)\to A(\lambda)\).

\subsection{Equation of motion for the adiabatic frame}

Differentiating \eqref{eq:WT} and using \eqref{eq:lambda_Schrodinger} we obtain
\begin{align*}
\frac{d}{d\lambda}W_T(\lambda)
 &= \dot A^\dagger(\lambda)U_T(\lambda)
    + A^\dagger(\lambda)\dot U_T(\lambda) \\
 &= \dot A^\dagger(\lambda)A(\lambda)W_T(\lambda)
 \\ &\quad - iT\,A^\dagger(\lambda)H(\lambda)A(\lambda)W_T(\lambda).
\end{align*}
Multiplying by \(i\) we get
\begin{equation}
\begin{aligned}
i\,\frac{d}{d\lambda}W_T(\lambda)
  &= K_T(\lambda)\,W_T(\lambda),
\\
K_T(\lambda)
  &:= i\,\dot A^\dagger(\lambda)A(\lambda)
     + T\,A^\dagger(\lambda)H(\lambda)A(\lambda).
\end{aligned}
\label{eq:first_order_WT}
\end{equation}

In the fixed orthonormal basis \(\{\ket{\varphi_n(0)}\}\), the
matrix elements of \(K_T(\lambda)\) are
\begin{equation*}
K_{mn}(\lambda)
 := \bra{\varphi_m(0)}K_T(\lambda)\ket{\varphi_n(0)}.
\end{equation*}
A straightforward computation using the explicit form of
\(\ket{\psi_n(\lambda;T)}\) shows that

(i) the diagonal entries vanish,
\begin{equation}
K_{nn}(\lambda)=0,
\label{eq:A6a}
\end{equation}

(ii) whereas for \(m\neq n\) one has
\begin{equation}
\begin{split}
K_{mn}(\lambda)
  ={}& i\,e^{iT[\alpha_m(\lambda)-\alpha_n(\lambda)]
           -i[\gamma_m(\lambda)-\gamma_n(\lambda)]}
  \\ &\times\braket{\dot\varphi_m(\lambda)}{\varphi_n(\lambda)}.
\end{split}
\label{eq:A6b}
\end{equation}
Thus \(K_T(\lambda)\) is purely off-diagonal in the instantaneous
eigenbasis, and its matrix elements are oscillatory
functions of \(T\).

\subsection{Integral equation and norm bound}

\rev{Equation~\eqref{eq:first_order_WT}} is a first-order linear differential
equation with solution
\begin{equation}
W_T(\lambda)
  = \mathbb{I} - i\int_0^\lambda
      K_T(\lambda')\,W_T(\lambda')\,d\lambda'.
\label{eq:WT_expansion}
\end{equation}
Introduce the operator
\begin{equation}
F_T(\lambda)
  := -i\int_0^\lambda K_T(\lambda')\,d\lambda'.
\end{equation}
An integration by parts using \eqref{eq:WT_expansion} gives
\begin{align}
W_T(\lambda)
 &= \mathbb{I}
    + F_T(\lambda)W_T(\lambda) \nonumber\\
 &\quad + i\int_0^\lambda
        F_T(\lambda')K_T(\lambda')W_T(\lambda')\,d\lambda'.
\label{eq:A9}
\end{align}

Let \(\|\cdot\|\) be any unitarily invariant submultiplicative
norm (for instance the operator norm).
Since \(W_T(\lambda)\) and \(A(\lambda)\) are unitary, we have
\(\|W_T(\lambda)\|=1\).
Taking norms in \eqref{eq:A9} and using
\(\|ABC\|\le\|A\|\|B\|\|C\|\) we obtain
\begin{align}
\|W_T(\lambda)-\mathbb{I}\|
 \le{}& \|F_T(\lambda)\| \nonumber\\
   &+ \int_0^\lambda
        \|F_T(\lambda')\|\,
        \|K_T(\lambda')\|\,d\lambda'.
\label{eq:A10}
\end{align}
By \eqref{eq:A6b} each matrix element of \(K_T(\lambda)\) has modulus
\(|\braket{\dot\varphi_m(\lambda)}{\varphi_n(\lambda)}|\),
which does not depend on \(T\); hence
\(\|K_T(\lambda)\|\) is bounded by a constant
independent of \(T\) and \(\lambda\).
Therefore, to prove \eqref{eq:what_we_want_to_prove} it suffices to show that
\begin{equation}
\sup_{\lambda\in[0,1]}\|F_T(\lambda)\| \xrightarrow[T\to\infty]{} 0.
\label{eq:FT}
\end{equation}

\subsection{Oscillatory integrals}

In the basis \(\{\ket{\varphi_n(0)}\}\),
\(F_T(\lambda)\) has matrix elements
\begin{equation}
(F_T)_{mn}(\lambda)
   = -i\int_0^\lambda K_{mn}(\lambda')\,d\lambda'.
\end{equation}
Using \eqref{eq:A6a}--\eqref{eq:A6b}, the diagonal entries vanish,
\((F_T)_{nn}(\lambda)=0\), and for \(m\neq n\),
\begin{equation}
(F_T)_{mn}(\lambda)
   = \int_0^\lambda
        e^{iT\theta_{mn}(\lambda')}\,
        f_{mn}(\lambda')\,d\lambda',
\label{eq:A13}
\end{equation}
where we defined
\begin{equation*}
\begin{aligned}
\theta_{mn}(\lambda)
  &:= \alpha_m(\lambda)-\alpha_n(\lambda)
     -\frac{1}{T}\bigl[\gamma_m(\lambda)-\gamma_n(\lambda)\bigr],
\\
f_{mn}(\lambda)
  &:= \braket{\dot\varphi_m(\lambda)}{\varphi_n(\lambda)}.
\end{aligned}
\end{equation*}
Note that \(\theta'_{mn}(\lambda)=\Delta_{mn}(\lambda)+O(1/T)\) and,
by the gap assumption, \(|\theta'_{mn}(\lambda)|\ge g/2>0\) for all
sufficiently large \(T\).
The functions \(f_{mn}\) and \(\theta'_{mn}\) are bounded and smooth
on \([0,1]\).

We now use integration by parts on \eqref{eq:A13}:
\begin{align}
(F_T)_{mn}(\lambda)
 &= \left[
     \frac{f_{mn}(\lambda')}{iT\,\theta'_{mn}(\lambda')}
     e^{iT\theta_{mn}(\lambda')}
    \right]_{\lambda'=0}^{\lambda'=\lambda}
    \nonumber\\
 &\quad - \int_0^\lambda
       e^{iT\theta_{mn}(\lambda')}
       \frac{d}{d\lambda'}
       \left(
         \frac{f_{mn}(\lambda')}{iT\,\theta'_{mn}(\lambda')}
       \right)
       d\lambda'.
\label{eq:A14}
\end{align}
Since \(|\theta'_{mn}(\lambda')|\ge g/2\) and
\(f_{mn},\theta'_{mn}\) and their derivatives are bounded,
both the boundary terms and the integral in \eqref{eq:A14} are
\(O(1/T)\).
Thus there exists a constant \(c>0\), independent of \(T\), such that
\begin{equation}
\sup_{\lambda\in[0,1]}
\bigl|(F_T)_{mn}(\lambda)\bigr|
  \le \frac{c}{T}
\quad\text{for all }m\neq n.
\end{equation}
Hence
\begin{equation}
\sup_{\lambda\in[0,1]}\|F_T(\lambda)\|
  \xrightarrow[T\to\infty]{} 0,
\label{eq:A16}
\end{equation}
which is precisely \eqref{eq:FT}.

Combining \eqref{eq:A10} and \eqref{eq:A16} yields
\begin{equation}
\sup_{\lambda\in[0,1]}
\|W_T(\lambda)-\mathbb{I}\|
  \xrightarrow[T\to\infty]{} 0.
\end{equation}
Recalling \eqref{eq:WT}, this proves the adiabatic theorem in the
form
\begin{equation}
\lim_{T\to\infty}
\left\|A^\dagger(\lambda)U_T(\lambda)-\mathbb{I}\right\| = 0,
\qquad\forall\,\lambda\in[0,1],
\end{equation}
\rev{i.e.,} in the slow-driving limit the true evolution \(U_T(\lambda)\)
coincides with the adiabatic intertwiner \(A(\lambda)\) and there are
no transitions between different instantaneous eigenspaces.

\subsection{Adiabatic condition for finite evolution time}

We now consider the case where the total evolution time $T$ is finite,
which is the relevant regime for constructing quantum gates.
Recall that in the instantaneous eigenbasis
$\{\ket{\varphi_k(\lambda)}\}$, the generator $K_T(\lambda)$ of the adiabatic
frame $W_T(\lambda)$ has purely off-diagonal matrix elements
\begin{multline*}
K_{kj}(\lambda) = i\,e^{iT[\alpha_k(\lambda)-\alpha_j(\lambda)]
             -i[\gamma_k(\lambda)-\gamma_j(\lambda)]}
             \\ \times\braket{\dot\varphi_k(\lambda)}{\varphi_j(\lambda)},
\qquad k\neq j,
\end{multline*}
where $\lambda\in[0,1]$ is the same rescaled time as above, the overdot denotes
$d/d\lambda$, $\alpha_n(\lambda)=\int_0^\lambda E_n(\lambda')\,d\lambda'$ are
dynamical phases, and $\gamma_n(\lambda)$ are the geometric phases defined
above.

If $K_{kj}(\lambda)\equiv0$ for all $k\neq j$ and all $\lambda$,
there is exactly no transition between different eigenspaces.
For finite $T$ this condition is far too strong.
Instead, we use the first-order Dyson approximation, which shows that
transitions are suppressed provided
\begin{equation}
F_{kj}(\lambda)
   := \int_0^\lambda K_{kj}(\lambda')\,d\lambda'
   \approx 0
   \quad \forall \lambda\in[0,1],\ \forall k\neq j.
\label{eq:Fkj_def}
\end{equation}

To obtain a clearer condition, write
\begin{equation}
K_{kj}(\lambda) = P_{kj}(\lambda)\,G_{kj}(\lambda),
\label{eq:PKG_def}
\end{equation}
with
\begin{equation}
\begin{aligned}
P_{kj}(\lambda) &:= e^{iT[\alpha_k(\lambda)-\alpha_j(\lambda)]},
\\
G_{kj}(\lambda) &:= i\,e^{-i[\gamma_k(\lambda)-\gamma_j(\lambda)]}\,
             \braket{\dot\varphi_k(\lambda)}{\varphi_j(\lambda)}.
\end{aligned}
\end{equation}
Thus $P_{kj}(\lambda)$ carries the rapidly oscillating dynamical phase,
while $G_{kj}(\lambda)$, which absorbs the slowly varying geometric phases,
is a slowly varying nonadiabatic coupling.

Define
\begin{equation}
Q_{kj}(\lambda) := \int_0^\lambda P_{kj}(\lambda')\,d\lambda'.
\label{eq:Q_def}
\end{equation}
Using $dQ_{kj}(\lambda) = P_{kj}(\lambda)\,d\lambda$ and \eqref{eq:PKG_def},
we can rewrite \eqref{eq:Fkj_def} as
\begin{align}
F_{kj}(\lambda)
 &= \int_0^\lambda K_{kj}(\lambda')\,d\lambda'
  = \int_0^\lambda G_{kj}(\lambda')\,P_{kj}(\lambda')\,d\lambda' \nonumber\\
 &= \int_0^\lambda G_{kj}(\lambda')\,dQ_{kj}(\lambda').
\label{eq:F_as_GdQ}
\end{align}

Applying integration by parts to \eqref{eq:F_as_GdQ} yields
\begin{align}
F_{kj}(\lambda)
 &= G_{kj}(\lambda)\,Q_{kj}(\lambda)
    - \int_0^\lambda Q_{kj}(\lambda')\,dG_{kj}(\lambda') \nonumber\\
 &= G_{kj}(\lambda)\,Q_{kj}(\lambda)
    - \int_0^\lambda Q_{kj}(\lambda')\,\frac{dG_{kj}}{d\lambda'}\,d\lambda'.
\label{eq:F_IBP}
\end{align}
(Here we used $Q_{kj}(0)=0$.)

Now take any unitarily invariant matrix norm $\|\cdot\|$.
From \eqref{eq:F_IBP} we obtain
\begin{align}
\|F_{kj}(\lambda)\|
 \le{}& \|G_{kj}(\lambda)\|\;\|Q_{kj}(\lambda)\| \nonumber\\
   &+ \int_0^\lambda \|Q_{kj}(\lambda')\|\;
              \left\|\frac{dG_{kj}}{d\lambda'}\right\|
              d\lambda'.
\label{eq:F_norm_bound}
\end{align}

Using the definition \eqref{eq:Q_def}, this condition depends only on
the dynamical phases:
\begin{equation}
\sup_{\lambda\in[0,1]}
\left|
\int_0^\lambda e^{iT[\alpha_k(\lambda')-\alpha_j(\lambda')]}\,d\lambda'
\right|
  \approx 0,
\quad \forall k\neq j.
\label{eq:finite_time_condition}
\end{equation}
Equation \eqref{eq:finite_time_condition} is our finite-time
adiabatic condition: if the oscillatory integrals of the dynamical
phase differences are uniformly small, then the first-order
off-diagonal amplitudes $F_{kj}(\lambda)$ are small, and eigenstate
transitions are suppressed to first order in the Dyson expansion.
%%%%%%%%%%%%%%%%%%%%%%%%%%%%%%%%%%%

\section{Rate independence of $\delta$-kick protocols}
\label{supp:theorem1_proof}
We now state and show a strengthened many-body version of the rate-independent plateau result, with the locality, boundedness, and gap assumptions made explicit.

\textit{Statement (rate-independent defect plateau for $\delta$-kick protocols in a gapped local phase).}

Let $\Lambda \subset \mathbb{Z}^d$ be a finite region with Hilbert space $\mathcal{H}_\Lambda$. Let
\begin{equation}
\label{eq:gapped_H}
H_\Lambda(s)=\sum_{X\subset \Lambda} h_X(s),
\qquad s\in[0,1],
\end{equation}
be a local Hamiltonian path with:

\begin{enumerate}
    \item Finite range / locality. Each $h_X(s)$ has support diameter
    \begin{equation}
    \operatorname{diam}(\operatorname{supp} h_X(s)) \le r_0,
    \qquad
    \sup_{s\in[0,1]} \|h_X(s)\| \le J .
    \end{equation}

    \item Uniform gap. For every $s\in[0,1]$ and all $\Lambda$, the spectrum has a ground-space projector $P_\Lambda(s)$ separated by a gap $\Delta>0$ from the rest (allowing finite degeneracy):
    \begin{equation}
    \operatorname{spec}(H_\Lambda(s))\cap (E_0(s),E_0(s)+\Delta)=\varnothing .
    \end{equation}

    \item Defect density observable. Assume the defect operator is a sum of bounded local terms:
    \begin{equation}
    \label{eq:defect_observable}D_\Lambda=\sum_{x\in\Lambda} d_x,
    \end{equation}
    with
    \begin{equation}
    \label{eq:bounded_local_terms}
    \operatorname{supp}(d_x)\subset B_\ell(x),
    \qquad
    \|d_x\|\le 1 .
    \end{equation}
\end{enumerate}
Here, $B_\ell(x)$ is the ball of radius $\ell$ centered at the lattice site $x$.
Define the defect density by
\begin{equation}
n_\Lambda(\psi):=\frac{1}{|\Lambda|}\,\langle \psi|D_\Lambda|\psi\rangle .
\end{equation}

Choose a gapped local Hamiltonian path $H_\Lambda(s)$ with $s\in[0,1]$. Apply $R$ $\delta$-kicks at fixed scaled times $s_j$ (for example, at the midpoints $s_j=\frac{2j-1}{2R}$) with fixed kick area $a$:
\begin{equation}
\label{eq:H_delta_jump}
H_\Lambda^{(T)}(t)=a\sum_{j=1}^{R}\delta(t-s_jT)\,H_\Lambda(s_j).
\end{equation}
Then for any total duration $T$, the evolution operator is
\begin{equation}
\label{eq:U_product}
U_{\Lambda,R}(T)=\prod_{j=1}^{R} e^{-ia H_\Lambda(s_j)},
\end{equation}
and hence is exactly independent of $T$.

Consequently, if the initial state $|\psi_0\rangle$ lies in the ground space at $s=0$, \rev{$|\psi_0\rangle\in\mathrm{Ran}\,P_\Lambda(0)$,}
and the final state is
\[
|\psi_{\Lambda,R}\rangle:=U_{\Lambda,R}(T)|\psi_0\rangle,
\]
then the post-protocol defect density
\[
n_{\Lambda,R}(T):=\frac{1}{|\Lambda|}\,
\langle \psi_{\Lambda,R}|D_\Lambda|\psi_{\Lambda,R}\rangle
\]
is exactly independent of $T$ for every finite $\Lambda$.

The key point is that the protocol Hamiltonian in \rev{Eq.~\eqref{eq:H_delta_jump}} has support only at the kick times $t=s_jT$. Hence, the time-ordered exponential factorizes exactly across the jumps:
\[
U_{\Lambda,R}(T)
=
\mathcal{T}\exp\!\left(
-i\int_0^T H_\Lambda^{(T)}(t)\,dt
\right)
=
\prod_{j=1}^{R} e^{-iaH_\Lambda(s_j)}.
\]
This is \rev{Eq.~\eqref{eq:U_product}}. Since no factor of $T$ appears anywhere in the ordered product, the full unitary is exactly independent of the total duration $T$. This already gives the plateau statement at the algebraic level.

\rev{For the final state $|\psi_{\Lambda,R}\rangle=U_{\Lambda,R}(T)|\psi_0\rangle$ defined above, the defect density is}
\[
n_{\Lambda,R}(T)
=
\frac{1}{|\Lambda|}
\langle \psi_0|
U_{\Lambda,R}(T)^\dagger D_\Lambda U_{\Lambda,R}(T)
|\psi_0\rangle .
\]
Using \rev{Eq.~\eqref{eq:U_product}}, this becomes
\begin{multline*}
n_{\Lambda,R}(T)
=
\frac{1}{|\Lambda|}
\langle \psi_0|
\left(\prod_{j=R}^{1} e^{iaH_\Lambda(s_j)}\right)
D_\Lambda
\\ \times
\left(\prod_{j=1}^{R} e^{-iaH_\Lambda(s_j)}\right)
|\psi_0\rangle ,
\end{multline*}
which contains no $T$. Therefore
\[
n_{\Lambda,R}(T)=n_{\Lambda,R}
\]
is exactly independent of $T$ for every finite system. $\blacksquare$

\begin{remark}
Assumptions 1--3 above are not used in the argument for $T$-independence, which is purely algebraic and holds for arbitrary Hamiltonian paths, initial states, and bounded observables. Their role is to delimit the physically relevant regime, gapped local many-body phases with local defect observables, in which the plateau value established in Appendix~\ref{supp:driven_bogoliubov} is small and locally determined.
\end{remark}

\section{Geodesics}
\label{supp:geodesics}
This appendix reviews geodesics and the associated quantum speed limit.

The Fubini--Study (FS) metric is the natural Riemannian metric on the projective Hilbert space $\mathbb{P}(\mathcal{H})$, where physical pure states live as rays $\lvert\psi\rangle \sim e^{i\phi}\lvert\psi\rangle$. It measures how distinguishable two nearby rays are while factoring out overall phase and normalization.
Given a state $\ket{\psi}$ in $\mathcal{H}$, the vectors that differ from it by a global phase are identified with a single point of $\mathbb{P}(\mathcal{H})$.

For a normalized state $\lvert\psi(\vec{\lambda})\rangle$, the line element is given by
\begin{eqnarray}
ds^2
&=& \langle d\psi \lvert d\psi\rangle - \lvert\langle \psi \lvert d\psi\rangle\rvert^2
= g_{ij}\, d\lambda^i d\lambda^j,
\\
g_{ij}&=&\mathrm{Re}\!\left[\langle \partial_i \psi \lvert (1-\lvert\psi\rangle\langle\psi\rvert)\rvert \partial_j \psi\rangle\right].
\end{eqnarray}
Here, $\partial_i$ refers to $\partial/\partial \lambda_i$.
At finite separation, the induced geodesic distance between pure states is
\begin{equation}
d_{\mathrm{FS}}(\psi,\phi)=\arccos\bigl\lvert\langle \psi \lvert \phi \rangle\bigr\rvert,
\end{equation}
i.e., \rev{half} the Bloch-sphere angle in the case \rev{of} qubits.
Operationally, the FS metric is the real part of the quantum geometric tensor $Q_{ij}=\langle \partial_i \psi \lvert (1-\lvert\psi\rangle\langle\psi\rvert)\rvert \partial_j \psi\rangle$; its imaginary part gives the Berry curvature:
\begin{equation}
\mathrm{Re}\,Q_{ij}=g_{ij},\quad
\mathrm{Im}\,Q_{ij}=\tfrac{1}{2}\,\mathcal{F}_{ij}.
\end{equation}
This ties state distinguishability to geometric phase. The FS arc length also controls quantum speed limits: the Anandan--Aharonov relation bounds the time to evolve from $\lvert\psi_0\rangle$ to $\lvert\psi_T\rangle$ by the FS geodesic length,
\begin{equation}
T \ge \frac{d_{\mathrm{FS}}(\psi_0,\psi_T)}{\overline{v}},\qquad
\overline{v}=\rev{\frac{\overline{\Delta H}}{\hbar}},
\end{equation}
where the ``speed'' is set by the time-averaged energy uncertainty $\overline{\Delta H}$, and $\overline{v}$ is the time-averaged speed. Thus, geodesics in the FS geometry represent time-optimal paths when $\Delta H$ is constrained.

In applications, $g_{ij}$ equals one quarter of the quantum Fisher information for pure states, $F_{ij}=4g_{ij}$, so FS geometry underlies quantum metrology and the Cram\'er--Rao bound \cite{Garcia-Pintos2022Feb}. In many-body systems, the FS metric is the fidelity susceptibility: it often peaks or diverges near quantum critical points, offering a coordinate-free probe of phase transitions and finite-size/Kibble--Zurek scaling. In adiabatic control and band topology, the pair $(g_{ij},\mathcal{F}_{ij})$ quantifies how ``curved'' the eigenstate bundle is over parameter space, governing diabatic errors, optimal adiabatic schedules, and topological responses.
See Refs.~\cite{Braunstein1994May,Chruscinski,Hetenyi2023,Zhang2023Jul,Tomka2016Jun,Kolodrubetz2013May,Cafaro2022Oct} and references therein.

\section{Many-body XY model}
\label{supp:XYmodel}
%%%%%%%%%%%%%%
Suppose we have an $N$-qubit system with the following Hamiltonian:
\begin{equation}
    H = -\sum_{j=1}^N \left[\frac{1+\gamma}{2} X_j X_{j+1}+ \frac{1-\gamma}{2}Y_j Y_{j+1}+ hZ_j \right],
\label{eqappen:XY}
\end{equation}
where $X,Y,Z$ are the standard Pauli matrices, and $\gamma$ and $h$ are external control parameters. This system can be solved analytically\rev{~\cite{Lieb1961,Katsura1962,Barouch1970}} if one takes the periodic boundary condition, i.e., $A_{N+1}=A_1$, where $A=X,Y,Z$. Using the \rev{Jordan--Wigner} transformation, followed by the \rev{momentum-space} representation, the XY Hamiltonian can be written as follows
\begin{equation}\label{eq:Hk-matrix}
H_k = -2\begin{pmatrix}
h - \cos k & \gamma \sin k \\
\gamma \sin k & -(h - \cos k)
\end{pmatrix},
\end{equation}
where
\begin{equation}\label{eq:ak-deltak}
a_k = h - \cos k,
\qquad
\Delta_k = \gamma \sin k.
\end{equation}

A straightforward Hamiltonian diagonalization leads to
\begin{equation}
\lambda_{\pm} = \pm 2\varepsilon_k,
\qquad
\varepsilon_k = \sqrt{a_k^2 + \Delta_k^2}.
\end{equation}

Let us compute the ground state $|\rev{\mathrm{GS}}\rangle = \alpha \ket{\uparrow _k}+\beta \ket{\downarrow _k}$. From the eigenvalue equation one finds
\begin{equation}\label{eq:alpha-beta}
\alpha = \frac{1}{\Delta_k}(a_k + \varepsilon_k)\beta.
\end{equation}

Defining the mixing angle $\tan \theta_k = \Delta_k/a_k$, the ground state reads
\begin{equation}\label{eq:groundstate0}
|\psi_0(\gamma,h)\rangle = \cos \big(\tfrac{\theta_k}{2}\big)\, |{\uparrow}_k\rangle
+ \sin \big(\tfrac{\theta_k}{2}\big)\, |{\downarrow}_k\rangle,
\end{equation}
where $|{\uparrow}_k\rangle \equiv (1,0)^T$ and $|{\downarrow}_k\rangle \equiv (0,1)^T$.

\subsection{Fubini--Study metric and geodesic trajectory}
\label{supp:Fubini--Study}
The geo and geo-jump strategies require us to compute the Fubini--Study metric \cite{Provost1980}.
Let us briefly discuss the Riemannian structures on manifolds of quantum states. Consider a closed many-body system described by the Hamiltonian $H(\vec{\lambda})$ which depends on a set of control parameters $\vec{\lambda}=(\lambda_1(t),\lambda_2(t), \rev{\ldots,} \lambda_p(t))$, where $p$ is the dimension of the parameter manifold $\mathcal{M}$. The FS metric endows the parameter manifold with a Riemannian structure, which can be recognized by addressing the problem of finding optimal trajectories on $\mathcal{M}$ and $\vec{\lambda}_{\rm opt}$ that maximize the instantaneous fidelity between two infinitesimally separated ground states
\begin{align}
    F=|\langle \psi_0(\vec{\lambda})|\psi_0(\vec{\lambda}+d\vec{\lambda})\rangle |^2.
\end{align}
One can also quantify the distance between those ground states in the Hilbert space as $ds^2=1-|\langle \psi_0(\vec{\lambda})|\psi_0(\vec{\lambda}+d\vec{\lambda})\rangle |^2=g_{\mu\nu}d\lambda^{\mu}d\lambda^{\nu}$, where the quantum metric tensor is given by~\cite{Provost1980}
\begin{align}
    g_{\mu\nu}={\rm Re}\big[\langle \partial_{\mu}\psi_0|\partial_{\nu}\psi_0\rangle- \langle \partial_{\mu}\psi_0|\psi_0\rangle \langle \psi_0|\partial_{\nu}\psi_0\rangle\big].
\end{align}
The expansion of $ds^2$ in $\{d\lambda^{\mu}\}$ shows that \rev{$g_{\mu\nu}$} induces a metric on $\mathcal{M}$; namely, $\mathcal{M}$ is a metric space that provides us with the notion of geodesic curves. On a Riemannian manifold, a geodesic is a path that minimizes the distance functional
\begin{align}
    \mathcal{L}(\vec{\lambda})=\int_{\vec{\lambda}_i}^{{\vec{\lambda}_f}}ds=\int_{0}^{t_f}\sqrt{g_{\mu\nu}\dot{\lambda}^{\mu}\dot{\lambda}^{\nu}}\,dt,
\end{align}
between two points $\vec{\lambda}_i=\vec{\lambda}(0)$ and $\vec{\lambda}_f=\vec{\lambda}(t_f)$, and $\dot{\lambda}^{\mu}=d\lambda^\mu/dt$. As the distance $\mathcal{L}$ is independent of the parametrization, we can choose \rev{$g_{\mu\nu}\dot{\lambda}^{\mu}\dot{\lambda}^{\nu}$} to be a constant as the geodesic protocol. This choice does not change $\mathcal{L}$; it fixes the parametrization. Applying the Cauchy--Schwarz inequality
\[
\Big(\int_0^{t_f} f\,h\,dt\Big)^2\le\int_0^{t_f}f^2\,dt\int_0^{t_f}h^2\,dt
\]
to the constant function $f=1$ and to the integrand of $\mathcal{L}$, $h=\sqrt{g_{\mu\nu}\dot{\lambda}^{\mu}\dot{\lambda}^{\nu}}$, one obtains
\[
\mathcal{L}^2\le t_f\int_0^{t_f}g_{\mu\nu}\dot{\lambda}^{\mu}\dot{\lambda}^{\nu}\,dt,
\]
with equality if and only if $h$ is proportional to $f=1$, i.e., if $g_{\mu\nu}\dot{\lambda}^{\mu}\dot{\lambda}^{\nu}$ is constant. Therefore, the constant-speed parametrization is the affine parametrization of the length-minimizing path \cite{Tomka2016Jun}. For the one-dimensional manifolds considered below, all monotonic paths between $\lambda_i$ and $\lambda_f$ have the same length, and the geodesic protocol amounts to traversing this path at constant Fubini--Study speed. In what follows, we apply the latter to the specific case of the many-body XY model.

During our investigation, we deal with a one-dimensional manifold $\mathcal{M}$ characterized by a single parameter $\lambda$, which can be a variable magnetic field $h(t)$ in the quantum Ising model or the anisotropy $\gamma(t)$ in the XY model. For the latter, the FS metric reads
\begin{equation}
\begin{split}
    g^{k}_{\gamma\gamma}={\rm Re}\big[&\langle \partial_\gamma\psi_0(\gamma,h)|\partial_\gamma\psi_0(\gamma,h)\rangle\\
    &-|\langle \partial_\gamma\psi_0(\gamma,h)|\psi_0(\gamma,h)\rangle|^2\big].
\end{split}
\end{equation}

Using the ground state in Eq.~\eqref{eq:groundstate0}, the Fubini--Study metric is
\begin{equation}\label{eq:gdelta}
g^{k}_{\gamma\gamma} = \frac{1}{4} \left(\frac{\partial \theta_k}{\partial \gamma}\right)^2,
\end{equation}
where
\begin{equation}\label{eq:dtheta-delta}
\frac{\partial \theta_k}{\partial \gamma}
= \frac{\sin k}{a_k}\cos^2 \theta_k.
\end{equation}
Since the mixing angle $\theta_k$ is defined through
\begin{equation}\label{eq:tan-theta-delta}
\tan \theta_k = \frac{\gamma \sin k}{a_k},
\end{equation}
one can readily compute its derivative with respect to $\gamma$. The latter gives us
\begin{equation}\label{eq:dtheta}
\frac{\partial \theta_k}{\partial \gamma}
= \frac{\sin k}{a_k} \cos^2 \theta_k,
\end{equation}
consistent with Eq.~\eqref{eq:dtheta-delta}.

On the other hand, the geodesic trajectory obeys the equation \cite{Tomka2016Jun}
\begin{equation}\label{eq:geodesic}
g^{k}_{\gamma\gamma}\, \dot{\gamma}^2 = \alpha^2 = \text{const}.\,\,{\rm or }\,\, \dot{\gamma} = 2\alpha/\big({\tfrac{\partial \theta_k}{\partial \gamma}}\big).
\end{equation}

Combining with Eq.~\eqref{eq:dtheta-delta}, and differentiating Eq.~\eqref{eq:tan-theta-delta} with respect to time, one finds
\begin{equation}\label{eq:thetadot}
\rev{\dot{\theta}_k} = 2\alpha.
\end{equation}

Integration yields
\begin{equation}\label{eq:theta-sol}
\theta_k(t) = \theta_{k,i} + \frac{\theta_{k,f} - \theta_{k,i}}{t_f - t_i}(t - t_i).
\end{equation}
Finally, the geodesic trajectory for $\gamma(t)$ reads
\begin{equation}\label{eq:delta-sol}
\gamma(t) = \frac{(h - \cos k)}{\sin k}\,\tan \theta_k(t).
\end{equation}

\subsection{Schedule shared by all modes}
\label{supp:global_schedule}
Equation~\eqref{eq:delta-sol} depends on $k$ through $a_k$ and the mixing angles, so it prescribes a different control for every mode. The ground state of the chain is the product of the mode ground states over $k=(2m-1)\pi/N$, $m=1,\dots,N/2$, and its quantum metric along the control line is the sum of the mode metrics, Eq.~\eqref{eq:global_metric}, with $g^{k}_{\lambda\lambda}=\frac{1}{4}(\partial\theta_k/\partial\lambda)^2$ and
\begin{equation}
\frac{\partial\theta_k}{\partial\gamma}=\frac{a_k\sin k}{a_k^2+\gamma^2\sin^2k},
\qquad
\frac{\partial\theta_k}{\partial h}=-\frac{\gamma\sin k}{a_k^2+\gamma^2\sin^2k}.
\label{eq:dtheta_both}
\end{equation}
The constant-speed condition $\sqrt{g_{\lambda\lambda}}\,|\dot\lambda|=\mathrm{const.}$ gives the schedule as the inverse of the normalized arc length,
\begin{equation}
\frac{t}{T}=\frac{\displaystyle\int_{\lambda_i}^{\lambda(t)}\sqrt{g_{\lambda'\lambda'}}\,d\lambda'}{\displaystyle\int_{\lambda_i}^{\lambda_f}\sqrt{g_{\lambda'\lambda'}}\,d\lambda'}.
\label{eq:global_schedule}
\end{equation}
In the thermodynamic limit the sum over the $N/2$ modes becomes $\frac{N}{2\pi}\int_0^{\pi}dk$, and the integrals have closed forms. On the Ising line, $\gamma=1$,
\begin{equation}
\frac{g_{hh}}{N}=\begin{cases}\dfrac{1}{16h^{2}(h^{2}-1)}, & h>1,\\[8pt] \dfrac{1}{16(1-h^{2})}, & h<1,\end{cases}
\label{eq:ghh_closed}
\end{equation}
which near $h=1$ is $1/(32|h-1|)$, the fidelity susceptibility of the transverse-field Ising chain. The arc length $\int\sqrt{g_{hh}}\,dh$ is proportional to $\arccos(1/h)$ for $h\ge1$ and to $-\arccos h$ for $h\le1$, and inverting Eq.~\eqref{eq:global_schedule} gives $h(s)$ of Eq.~\eqref{eq:closed_schedules}, with $L=\arccos(1/h_i)+\pi/2$ for $h_i>1$ and $h_f=0$. On the anisotropy and gapless lines the metric for the ramp of $\gamma$ at fixed $h$ does not depend on $h$ for $|h|\le1$,
\begin{equation}
\frac{g_{\gamma\gamma}}{N}=\frac{1}{16|\gamma|(1+|\gamma|)^{2}},
\label{eq:ggg_closed}
\end{equation}
so the arc length is proportional to $\mathrm{sgn}(\gamma)\arctan\sqrt{|\gamma|}$ and the schedule for $\gamma\colon-1\to1$ is $\gamma(s)$ of Eq.~\eqref{eq:closed_schedules}. We checked both closed forms against the sum of Eq.~\eqref{eq:global_metric} with $N=4\times10^{5}$ to machine precision. In both cases $g\propto1/|\lambda-\lambda_c|$ at the critical point, so $\sqrt{g}\,d\lambda$ is proportional to $d\sqrt{|\lambda-\lambda_c|}$ and the schedule crosses the critical point as $|\lambda-\lambda_c|\propto(s-s_c)^2$. At $s=0.1$, $0.25$, $0.5$, $0.75$, and $0.9$ the Ising-line schedule takes the values $h=2.54$, $1.32$, $1.00$, $0.69$, and $0.30$, and the $\gamma$ schedule takes $\gamma=-0.528$, $-0.172$, $0$, $0.172$, and $0.528$. The finite-$N$ sum of Eq.~\eqref{eq:global_metric} differs from these limits only within $|\lambda-\lambda_c|\lesssim1/N$, where it saturates; for $N=250$ the kick fields of the geo-jump protocol differ from the thermodynamic ones by up to $0.05$ on the Ising line, and the plateau of Fig.~\ref{fig:KZ_scaling_results}(c) by $0.001$ ($0.136$ against $0.137$).

\section{Kibble--Zurek mechanism (KZM)}
\label{supp:KZM}
\rev{The Kibble--Zurek mechanism (KZM)} describes the universal nonadiabatic dynamics of systems driven across a continuous phase transition at a finite rate \cite{Kibble1980Dec,Zurek1985Oct,Schutzhold2006,Uhlmann2007,Fischer2008,Uhlmann2010,Uhlmann2010b,Polkovnikov2011Aug,Polkovnikov2005Oct,Bando2020Sep,Dziarmaga2010Nov,Tang2025,Grabarits2025b}. When a control parameter $\lambda(t)$ is tuned through a critical point $\lambda_c$, the system's relaxation time diverges as
\begin{equation}
    \tau \sim |\varepsilon|^{-z\nu}, \qquad
\varepsilon = \frac{\lambda - \lambda_c}{\lambda_c},\nonumber
\end{equation}
where $\nu$ and $z$ are the correlation-length and dynamical critical exponents, respectively. The divergence of $\tau$ near criticality causes a breakdown of adiabatic evolution, leading to a ``freeze-out'' of correlations and the emergence of universal nonequilibrium length and time scales.

For a general power-law quench of the form
$
    \varepsilon(t) = \mathrm{sgn}(t)\left|\frac{t}{\tau_Q}\right|^{r},
$
the crossover (freeze-out) time $\hat{t}$ is defined by the condition
$
  \tau(\hat{\varepsilon}) \sim \left|\frac{\varepsilon}{\dot{\varepsilon}}\right|_{t=\hat{t}}\,.
$
Solving this yields the characteristic KZ scaling relations,
\begin{equation}
    \hat{\xi} \sim \xi_0 \left( \frac{\tau_Q}{\tau_0} \right)^{\frac{\nu r}{1 + r z \nu}},
\qquad
\hat{\tau} \sim \tau_0 \left( \frac{\tau_Q}{\tau_0} \right)^{\frac{z \nu r}{1 + r z \nu}},
\end{equation}
which govern the universal properties of the system after the transition.
The density of topological defects or excitations generated during the quench scales as
\begin{equation}
   n_{\mathrm{defect}} \sim \hat{\xi}^{-(d-p)}
\;\propto\;
\tau_Q^{-\frac{\nu r (d-p)}{1 + r z \nu}},
\end{equation}
where $d$ is the spatial dimension and $p$ the dimensionality of the defect (e.g., $p=0$ for point defects, $p=1$ for line defects).

Therefore, KZM links nonequilibrium dynamics to equilibrium critical exponents, providing a universal framework applicable to both classical and quantum phase transitions. It predicts scaling relations for defect densities, residual energies, and excitation probabilities that depend only on $(\nu, z)$, the ramp protocol, and system dimensionality, but are independent of microscopic details.

Recent studies indicate that there exists a universal breakdown of KZM across a phase transition and that typically happens at fast quenches \cite{Chesler2015May,Zeng2023Feb}.
When a quantum system is embedded in an open quantum system environment, \rev{anti-Kibble--Zurek} scaling can be observed, i.e., the universal breakdown happens at slow quenches \cite{Dutta2016Aug,Puebla2020Jun,PhysRevB.111.104310}. Noisy quenches across a quantum critical point also have interesting consequences in entanglement production, with next-nearest concurrences \rev{scaling} logarithmically with the quench rate \cite{Jafari2025}.

For the three paths of Fig.~\ref{fig:KZ_scaling_results} the asymptotic exponents are $1/2$ for the linear ramp of $\gamma$ at fixed $h=0.5$ across the anisotropic transition~\cite{Mukherjee2007}, $1/3$ for the ramp of $\gamma$ along the gapless line $h=1$ through the multicritical point~\cite{Divakaran2008}, and $1/2$ for the transverse-field ramp of the Ising chain~\cite{Dziarmaga2005}. In the units of Eq.~\eqref{eq:HXY} the lin ramp of panel (c) has $\tau_Q=T/10$, and the exact solution of Ref.~\cite{Dziarmaga2005} holds, as stated there, for $\tau_Q\gg4/\pi^3\simeq0.13$; the window $0.1\le\nu\le0.8$ of Fig.~\ref{fig:KZ_scaling_results} corresponds to $0.13\le\tau_Q\le1$, so the fitted values $0.55$, $0.38$, and $0.46$ are effective exponents of the crossover. Figure~\ref{fig:slow} follows the local exponent $\alpha=-d\ln n_{ex}/d\ln T$ down to $\nu=10^{-3}$ in the thermodynamic limit: for the lin ramps it settles at $0.50$, $0.33$, and $0.50$ below $\nu\simeq5\times10^{-3}$, the asymptotic values. For a ramp that crosses the critical point as $|\lambda-\lambda_c|\propto|t-t_c|^{r}$ the exponent becomes $r\nu d/(1+rz\nu)$~\cite{Sen2008,Barankov2008}, where in this formula $\nu$ is the correlation-length exponent defined above, i.e., $2/3$ for $r=2$ and $d=z=\nu=1$. The shared geodesic schedule of Eq.~\eqref{eq:closed_schedules} is of this type, $|\lambda-\lambda_c|\propto(t-t_c)^2$ (Appendix~\ref{supp:global_schedule}). Its local exponent on the Ising line decreases from $0.81$ in the window of Fig.~\ref{fig:KZ_scaling_results}(c) to $0.668$ at $\nu=10^{-3}$ [Fig.~\ref{fig:slow}(b)], close to the value $2/3$ quoted in Table~\ref{table:summary}; on the anisotropy line, where \rev{$d=z=1$ and the correlation-length exponent is $1$} as well, it reaches $0.669$; on the gapless line, which passes through the multicritical point, it settles at $0.40$, consistent with $r/(2r+1)=2/5$ for a crossing with $r=2$ along that line (the same scaling argument gives $1/3$ for $r=1$~\cite{Divakaran2008}).

\section{Driven Bogoliubov mode with delta-like kicks}\label{supp:driven_bogoliubov}

The kick protocol below is the $\pi$ pulse sequence of Refs.~\cite{Wang2016May,Xu2019Jun,Xing2026}, applied here to each Bogoliubov mode of the many-body system. We choose the Ising line by setting $\gamma=1$ in the \rev{XY model}~\eqref{eqappen:XY}. Each Bogoliubov \rev{\(k\) mode} is driven with a train of
\(\delta\)-like kicks along a path \rev{parametrized} by an angle \(\theta\):
\[
\tan\theta_k (h) = \frac{h - \cos k}{\sin k},
\qquad
\theta_i = \theta_k(h_i),
\qquad
\theta_f = \theta_k(h_f),
\]
where $h(t)$ follows the schedule shared by all modes, Eq.~\eqref{eq:global_schedule}, so that $\theta_k(t)=\theta_k(h(t))$. For the mode-resolved schedule of Eq.~\eqref{eq:geo_schedules} one has instead $\theta_k(t)=\theta_i+(\theta_f-\theta_i)\,t/T$.
Note that this angle is the complement of the mixing angle used in Appendix~\ref{supp:XYmodel}: for $\gamma=1$ one has $\hat{\mathbf n}(\theta)=(\cos\theta,\,0,\,\sin\theta)$, so that $\theta$ is measured from the $\hat{\mathbf x}$-axis of the Bloch sphere.
At the kick times \(t_j\) (midpoints in the scaled variable \(\lambda = t/T\)), the field takes the value $h_j=h(t_j)$; for the mode-resolved schedule,
\begin{equation}
\label{eq:h_kick_schedule}
    h_j = \sin k\, \tan\theta_k (t_j) + \cos k.
\end{equation}
The corresponding \(k\)-mode Hamiltonian becomes
\begin{equation}
\begin{split}
 H_k(t)
&= -2\,J(t)\big[(h(t) - \cos k)\,\sigma_z
+ \sin k\,\sigma_x\big]
\\ &= -2\,J(t)\,E_k(\theta)\,
\hat{\mathbf{n}}(\theta)\!\cdot\!\boldsymbol{\sigma},
\end{split}
\end{equation}
where
\begin{equation}
    E_k(\theta) = \sin k \,\sqrt{\gamma^2 + \tan^2\theta},
\qquad
\hat{\mathbf{n}}(\theta) =
\frac{(\gamma,\,0,\,\tan\theta)}{\sqrt{\gamma^2 + \tan^2\theta}}.
\end{equation}
The drive envelope \(J(t)\) consists of a train of delta-function spikes:
\begin{equation}
    J(t)
= \frac{\pi}{2}\sum_{j=1}^{R}\delta(t - t_j),
\qquad
t_j = \lambda_j T,
\qquad
\lambda_j = \frac{2j-1}{2R},
\end{equation}
so that each kick has a total area
\begin{align}
\int J(t)\,dt = \frac{\pi}{2}.
\end{align}

\subsection{Exact single-mode evolution as a product of SU(2) rotations}

Between kicks, the envelope \(J(t) = 0\), so all the dynamics occur instantaneously at the kick times $t_j$.
The unitary operator for the \rev{\(j\)th} kick is
\[
\begin{split}
U_{k,j}
&= \rev{\mathcal{T}\exp\!\left(-i\!\int_{t_j^-}^{t_j^+} H_k(t')\,dt'\right)}
\\ &= \exp\!\big\{\,i\pi\,E_k(\theta_j)\,
\hat{\mathbf{n}}(\theta_j)\!\cdot\!\boldsymbol{\sigma}\big\}
\\ &= \cos\alpha_{kj}\,I
+ i\sin\alpha_{kj}\,
\hat{\mathbf{n}}(\theta_j)\!\cdot\!\boldsymbol{\sigma},
\end{split}
\]
where the kick angle is defined as
\[
\alpha_{kj} = \pi\,E_k(\theta_j)
= \pi\,\sin k\,\sqrt{\gamma^2 + \tan^2\theta_j}.
\]
The total mode-resolved unitary is the ordered product of these SU(2) rotations:
\[
U_k = \prod_{j=1}^{R} U_{k,j}.
\]

\noindent
Let \(\ket{g_i(k)}\) denote the ground-state spinor aligned with
\(\hat{\mathbf{n}}(\theta_i)\),
and \(\ket{g_f(k)}\) the ground-state spinor aligned with
\(\hat{\mathbf{n}}(\theta_f)\).
The excitation probability for mode \(k\) after the entire sequence of kicks is
\begin{equation}
    p_k = |\langle
{e_f(k) | U_k | g_i(k)\rangle}|^2,
\qquad
\ket{e_f(k)} =
\text{excited spinor}\rev{.}
\end{equation}
Finally, the total defect (kink) density is given by the standard integral
\begin{equation}
    n_{\mathrm{defect}}^{\rev{\mathrm{geo.jump}}}
= \frac{1}{\pi}
\int_{0}^{\pi} dk\, p_k
\;=\frac{1}{2\pi}\int_{-\pi}^{\pi} dk\,p_k.
\end{equation}

\noindent
We note that the expression above constitutes the exact analytical reduction of the
geo-jump protocol.
All quantities are explicitly determined by the sequence
\(\{\theta_j\}\) and the parameters
\((\gamma,h_i, h_f, R)\)
through the matrices \(U_{k,j}\) defined above.

\subsection{Co-moving frame: an exact bound and the first-order amplitude}
\label{supp:comoving}

Let $\ket{g_j}$ and $\ket{e_j}$ be the eigenstates of $\hat{\mathbf n}(\theta_j)\cdot\boldsymbol\sigma$ with eigenvalues $+1$ and $-1$, i.e., the instantaneous ground and excited states at the \rev{$j$th} kick. In this basis the kick is diagonal,
\begin{equation}
U_{k,j}=e^{i\alpha_{kj}}\ketbra{g_j}{g_j}+e^{-i\alpha_{kj}}\ketbra{e_j}{e_j},
\label{eq:kick_diag}
\end{equation}
so a kick never changes the weight of the state on $\ket{g_j}$. Transitions arise only from the rotation of the frame between kicks. The axes $\hat{\mathbf n}(\theta_j)$ and $\hat{\mathbf n}(\theta_{j+1})$ differ by the angle $\Delta_j=\theta_{j+1}-\theta_j$ about $\hat{\mathbf y}$, with $\Delta_0=\theta_1-\theta_i$ and $\Delta_R=\theta_f-\theta_R$ at the two ends, and
\begin{equation}
|\braket{g_{j+1}}{g_j}|=\cos\tfrac{\Delta_j}{2},\qquad |\braket{e_{j+1}}{g_j}|=\sin\tfrac{|\Delta_j|}{2}.
\label{eq:frame_overlaps}
\end{equation}

\paragraph{Exact bound.} The Fubini--Study distance $d(\psi,\phi)=\arccos|\braket{\psi}{\phi}|$ is invariant under unitaries and obeys the triangle inequality. Since $U_{k,j}\ket{g_j}\propto\ket{g_j}$, the \rev{$j$th} kick leaves $d(\psi,g_j)$ unchanged, and moving the reference state from $\ket{g_j}$ to $\ket{g_{j+1}}$ changes the distance by at most $d(g_j,g_{j+1})=|\Delta_j|/2$. Starting from $\ket{g_i}$ and adding up the steps,
\begin{equation}
d(U_k g_i,\,g_f)\le\frac{1}{2}\sum_{j=0}^{R}|\Delta_j|=\frac{\Theta_k}{2}
\end{equation}
for a monotonic ramp, where $\Theta_k=|\theta_f-\theta_i|$ is the angle between the initial and final fields of the mode. Hence
\begin{equation}
p_k\le\sin^2\frac{\Theta_k}{2}=p_k^{\mathrm{sudden}},
\qquad
n^{\mathrm{geo.jump}}_{\mathrm{defect}}\le n_{\mathrm{sudden}},
\label{eq:bound}
\end{equation}
for any number of kicks and any kick angles. The protocol never creates more defects than a sudden quench. For the ramp $h\colon10\to0$ one has $n_{\mathrm{sudden}}=0.475$.

\paragraph{First order in the frame rotation.} Keeping a single transition $g\to e$, at the step $m$, together with the phases of the kicks before and after it, Eqs.~\eqref{eq:kick_diag} and \eqref{eq:frame_overlaps} give, to first order in the $\Delta_j$ and up to an overall phase,
\begin{equation}
\braket{e_f}{U_k\,g_i}\simeq\sum_{m=0}^{R}\sin\frac{\Delta_m}{2}\,e^{2i\Phi_m},
\qquad
\Phi_m=\sum_{j\le m}\alpha_{kj},
\label{eq:first_order}
\end{equation}
with $\Phi_0=0$. Each term is a small rotation of the frame, weighted by the phase $2\Phi_m$ that the kicks have accumulated between the ground and excited components up to that point. The excitation is small when these phases make the sum cancel, which is the content of the generalized adiabatic condition of Sec.~\ref{sec:gac} for a kicked evolution. For the small ramps considered next, Eq.~\eqref{eq:first_order} reproduces the exact defect density to a relative accuracy of about $10^{-5}$, and the bound \eqref{eq:bound} is obeyed by the exact $p_k$ mode by mode.

\subsection{Small ramps: the plateau decreases as $1/R$}
\label{supp:small_ramp}

For a small ramp, $\Theta_k\ll1$, the kick angle $\alpha_{kj}\simeq\alpha_k$ is nearly the same for all kicks, and the midpoint kicks give $\Delta_0=\Delta_R=\Theta_k/2R$ and $\Delta_m=\Theta_k/R$ otherwise. The sum in Eq.~\eqref{eq:first_order} is then a geometric series with trapezoidal weights, and
\begin{equation}
p_k\simeq\left(\frac{\Theta_k}{2R}\right)^{2}\cot^{2}\alpha_k\,\sin^{2}(R\alpha_k).
\label{eq:pk_small}
\end{equation}
Away from the resonances $\alpha_k=l\pi$, $l$ an integer, this is $O(R^{-2})$. At a resonance the kick is $\pm I$ and does nothing to the mode, so the frame rotations add up coherently and $p_k$ has a peak of height $\Theta_k^2/4$ and width $\sim1/(R\,|d\alpha_k/dk|)$. Integrating over $k$,
\begin{equation}
n^{\mathrm{geo.jump}}_{\mathrm{defect}}=\frac{C}{R}+O(R^{-2}),
\qquad
C=\sum_{k^*}\frac{\Theta_{k^*}^{2}}{4\pi\,|dE_k/dk|_{k^*}},
\label{eq:CR}
\end{equation}
where the sum runs over the momenta $k^*$ at which $E_{k^*}$ is an integer. The plateau is therefore never a constant in $R$: it vanishes as $1/R$, and only its $T$-independence is exact. For the ramp $h\colon1.5\to1.48$ at $\gamma=1$, Eq.~\eqref{eq:CR} gives $C=1.68\times10^{-5}$, the exact mode product gives $Rn=1.68\times10^{-5}$ at $R=5000$ [Fig.~\ref{fig:nR}(b)], and the local exponent of $n(R)$ lies between $-0.999$ and $-1.001$ for $R\ge10^{3}$. The ramps $0.5\to0.48$ and $1.2\to1.18$ behave in the same way, with $Rn/C\to1$; for $3.0\to2.98$ the ratio is $0.98$, because two of its resonances start at the zone edges, where $\Theta_k^2/|dE_k/dk|$ varies on the scale over which the resonance moves during the ramp, and the local form of Eq.~\eqref{eq:CR} does not apply to them. For comparison, $n_{\mathrm{sudden}}=1.85\times10^{-5}$ for the ramp $h\colon1.5\to1.48$, so the plateau sits below the bound \eqref{eq:bound} for every $R$.

\subsection{The ramp $h\colon 10\to0$ and the number of kicks}
\label{supp:nR}

\begin{figure*}[t]
        \centering
        \includegraphics[width=1.0\linewidth]{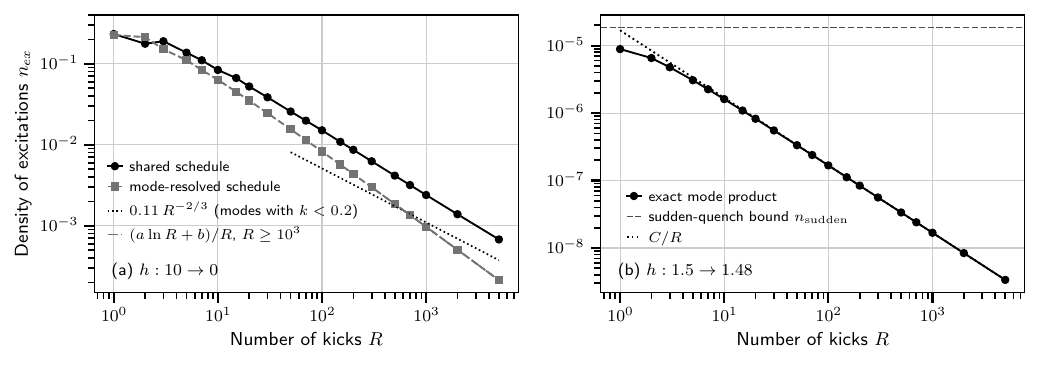}
        \caption{Defect density of the geo-jump protocol on the Ising line against the number of kicks $R$, from the exact mode product, Eq.~\eqref{eq:exact_nk}, on a fine momentum grid. (a) The ramp of Fig.~\ref{fig:KZ_scaling_results}(c), $h\colon10\to0$, for the schedule shared by all modes, Eq.~\eqref{eq:closed_schedules} (circles), and for the mode-resolved schedule (squares); the dotted lines are $0.11\,R^{-2/3}$, the contribution of the modes with $k<0.2$ (Appendix~\ref{supp:nR}), and $(a\ln R+b)/R$ with $a=0.059$, $b=0.56$, fitted for $R\ge10^{3}$. (b) The small ramp $h\colon1.5\to1.48$ with the mode-resolved schedule (for a ramp this small the kick fields of the two schedules differ by less than $2\times10^{-4}$ and the defect densities by a relative $10^{-5}$): exact values (circles), the $C/R$ law of Eq.~\eqref{eq:CR} (dotted), and the sudden-quench bound of Eq.~\eqref{eq:bound} (dashed).}
        \label{fig:nR}
\end{figure*}

The quench of Fig.~\ref{fig:KZ_scaling_results}(c) is not a small ramp, and the mode product must be evaluated as it stands. Figure~\ref{fig:nR}(a) shows $n(R)$ for the shared schedule of Eq.~\eqref{eq:closed_schedules} and for the mode-resolved schedule. With $R=5$ kicks, the values on the momenta of the $N=250$ chain are $0.137$ and $0.111$; at $R=1$ they are $0.233$ and $0.227$, against $n_{\mathrm{sudden}}=0.475$, and at $R=10$ they are $0.084$ and $0.063$. Beyond a few kicks $n(R)$ decreases, and for $R\gg1$ the two schedules differ. With the mode-resolved schedule the decrease follows $(a\ln R+b)/R$ with $a\simeq0.059$ and $b\simeq0.56$, fitted for $R\ge10^{3}$; the logarithm comes from the end point $h_f=0$, where $E_k=1$ for every $k$ at $\gamma=1$, so that all modes finish on the resonance $\alpha_k=\pi$. With the shared schedule, the modes with small $k$ ($k<0.2$ in Fig.~\ref{fig:nR}) have small kick angles near $h=1$, so for them the $R$ kicks act as Trotter steps of a continuous ramp of duration $\pi R/2$, and they reproduce the Kibble--Zurek law of that ramp with $\tau_Q\to\pi R/2$. The crossing of $h=1$ is quadratic, Eq.~\eqref{eq:closed_schedules}, so these modes follow the nonlinear-crossing law of Appendix~\ref{supp:KZM}: the modes with $k<0.2$ contribute $0.11\,R^{-2/3}$, with a local exponent of $-0.668$ from $R=500$ to $5000$, the exponent of the geo row of Table~\ref{table:summary}. The modes away from the critical point contribute $(0.12\ln R+0.41)/R$, fitted for $R\ge200$, with the same logarithm as the mode-resolved schedule, and they still make up $43\%$ of the total at $R=5000$. The local exponent of the total is therefore $-0.78$ at $R=5000$, and it approaches $-2/3$ only beyond the range of Fig.~\ref{fig:nR}.

\subsection{Cross-check: continuous ramp}

If, instead, the drive envelope is continuous, \(J(t)\equiv 1\), and
\(\theta(t)\) varies smoothly, the adiabatic-frame Hamiltonian acquires the standard
nonadiabatic term
\[
H_{\mathrm{ad}}(t)
= -\frac{1}{2}\,\dot{\theta}(t)\,\sigma_y.
\]
In this case, the usual Kibble--Zurek scaling for the transverse-field Ising model
is recovered,
\[
n_{\mathrm{defect}} \sim \tau_Q^{-1/2},
\]
for a linear crossing of the critical point, and $\tau_Q^{-2/3}$ for the quadratic crossing of Eq.~\eqref{eq:closed_schedules} (Table~\ref{table:summary}, Appendix~\ref{supp:KZM}, and Fig.~\ref{fig:slow}).
In the kick protocol the nonadiabatic coupling acts only through the frame rotations between kicks, Eq.~\eqref{eq:first_order}; the quench time drops out and the number of kicks takes its place (Appendix~\ref{supp:nR}).

\subsection{Exact expression (for all \(R\))}
The exact expression for the defect (kink) density is
\begin{multline}
\label{eq:exact_nk}
    n_{\mathrm{defect}}^{\textrm{geo.jump}}
= \frac{1}{\pi}
\int_{0}^{\pi} dk\,
\Bigg|\Bigg\langle e_f(k)\,\Bigg|
\\ \times
\prod_{j=1}^{R}
\Big[
\cos\alpha_{kj}
+ i\sin\alpha_{kj}\,
\hat{\mathbf{n}}(\theta_j)\!\cdot\!\boldsymbol{\sigma}
\Big]
\,\Bigg|\,g_i(k)\Bigg\rangle
\Bigg|^2,
\end{multline}
where
\[
\begin{aligned}
\alpha_{kj} &=
\pi\,\sin k\,\sqrt{\gamma^2 + \tan^2\theta_j},
\\
\hat{\mathbf{n}}(\theta_j)
&=
\frac{(\gamma,\,0,\,\tan\theta_j)}
{\sqrt{\gamma^2 + \tan^2\theta_j}},
\\
\theta_j
&=
\theta_k(h_j),
\end{aligned}
\]
with $h_j$ the field at the \rev{$j$th} kick: $h_j=h(\lambda_j)$ for the shared schedule, Eq.~\eqref{eq:global_schedule}, and $h_j=\sin k\tan[\theta_i+(j-\tfrac12)(\theta_f-\theta_i)/R]+\cos k$ for the mode-resolved one. The large-$R$ behavior of Eq.~\eqref{eq:exact_nk} is given by Eq.~\eqref{eq:CR} for small ramps and is shown in Fig.~\ref{fig:nR} for the ramp of Fig.~\ref{fig:KZ_scaling_results}(c).

\section{Topologically ordered system with degenerate ground space (toric code)}
\label{supp:toric_code}

Here the key new ingredient is ground-space degeneracy together with local indistinguishability\rev{. For a commuting-projector Hamiltonian the gap is stable under weak local perturbations if the ground states are locally indistinguishable and the local ground spaces are consistent with the global one~\cite{Bravyi2010}; the toric code meets both conditions. We first recall the definition of the toric code~\cite{Kitaev2003,Dennis2002} and the properties that are used below}.

\paragraph{Model: toric code (commuting-projector Hamiltonian).}
\leavevmode{}\rev{Consider an $L\times L$ square lattice ($L\ge2$) with periodic boundary conditions in both directions (a torus) and one qubit on every edge. There are $L^2$ vertices $v$, $L^2$ plaquettes $p$, and $2L^2$ qubits. To every vertex and every plaquette one assigns the star and plaquette operators
\begin{equation}
A_v=\prod_{e\ni v}X_e,
\qquad
B_p=\prod_{e\in\partial p}Z_e,
\label{eq:star-plaquette}
\end{equation}
the product of the Pauli operators $X_e$ on the four edges that meet at $v$ and the product of $Z_e$ on the four edges around $p$. Each of them squares to the identity and has the eigenvalues $\pm1$. A star and a plaquette share either no edge or two edges, so all the $A_v$ and $B_p$ commute with one another. The Hamiltonian is}
\begin{equation}
H_{\mathrm{TC}}
=
- J_e \sum_v A_v
- J_m \sum_p B_p ,
\label{eq:TC-Hamiltonian}
\end{equation}
\rev{with $J_e,J_m>0$. Up to a constant it is a sum, with positive weights, of the commuting projectors $(1-A_v)/2$ and $(1-B_p)/2$, which is why it is called a commuting-projector Hamiltonian; the $A_v$ and $B_p$ are the stabilizers of the corresponding quantum code.}

\paragraph{\rev{Ground space.}}
\rev{The ground states are the states with $A_v=B_p=+1$ for all $v$ and $p$, and their energy is $E_0=-(J_e+J_m)L^2$. On the torus every edge belongs to two stars and to two plaquettes, so that $\prod_vA_v=\prod_pB_p=1$. These are the only relations among the operators. Hence $2L^2-2$ of them are independent, and the ground space has dimension $2^{2L^2-(2L^2-2)}=4$. A basis of it is labeled by the eigenvalues of the products of $Z_e$ along the two loops that wind around the torus. An operator supported on a region that does not wrap around the torus has the same expectation value in every ground state and no matrix element between different ones; this is the local indistinguishability mentioned above.}

\paragraph{\rev{Excitations.}}
\rev{A vertex with $A_v=-1$ carries an electric charge $e$, and a plaquette with $B_p=-1$ carries a magnetic vortex $m$. With $N_e$ and $N_m$ the numbers of the two kinds of particles, $H_{\mathrm{TC}}=E_0+2J_eN_e+2J_mN_m$. These pointlike excitations are Abelian anyons: the wavefunction acquires a factor $-1$ when an $e$ particle is carried around an $m$ particle. Because of the two product relations each kind appears in pairs. The product of $Z_e$ along an open path of the lattice (a single edge in the simplest case) creates two $e$ particles at its end vertices, and the product of $X_e$ along an open path of the dual lattice creates two $m$ particles at its end plaquettes. The gap above the ground space is therefore $4\min(J_e,J_m)$. The anyons play the role that the kinks play in the Ising chain: they are the defects that a nonadiabatic evolution leaves behind.}

\paragraph{\rev{Defect density.}}
A natural ``defect operator'' is the anyon \rev{number} operator
\begin{equation}
D
=
\sum_v \frac{1-A_v}{2}
+
\sum_p \frac{1-B_p}{2},
\label{eq:defect-operator}
\end{equation}
\rev{in which $(1-A_v)/2$ and $(1-B_p)/2$ project onto the presence of an $e$ particle at $v$ and of an $m$ particle at $p$, so that $D=N_e+N_m$. The} corresponding defect density is
\begin{equation}
n
=
\frac{1}{|\Lambda|}\,\langle D \rangle \rev{,}
\label{eq:defect-density}
\end{equation}
\rev{where $|\Lambda|=2L^2$ is the number of vertices plus plaquettes, so that $0\le n\le1$.}

In any ground state,
\begin{equation}
\langle A_v \rangle = \langle B_p \rangle = 1,
\label{eq:ground-state-stabilizers}
\end{equation}
hence
\begin{equation}
n = 0.
\label{eq:n-zero-ground-state}
\end{equation}

\paragraph{\rev{Exact plateau $n=0$} for the commuting-path geo-jump protocol (explicit calculation).}
Take a gapped path that stays within the commuting-projector family, \rev{with the operators $A_v$ and $B_p$ fixed and only the couplings varying}:
\begin{equation}
H(s)
=
- J_e(s)\sum_v A_v
- J_m(s)\sum_p B_p \rev{,}
\label{eq:path-Hs}
\end{equation}
\rev{with $J_e(s),J_m(s)>0$ for all $s\in[0,1]$.} Then for any $s,s'$,
\begin{equation}
[H(s),H(s')] = 0,
\label{eq:Hs-commute}
\end{equation}
because all $A_v$ and $B_p$ commute.
We emphasize that the result below is established for this commuting-projector family, along which the ground space is $s$-independent; paths that deform the ground space (e.g., the toric code in an external field) lie outside the scope of the present calculation and are left for future work. \rev{Since the ground space does not move, the Fubini--Study metric vanishes along this path; there is no geodesic schedule to choose, and the kick times $s_j$ are arbitrary.}

\paragraph{Step 1: collapse the kick product.}
Define
\begin{equation}
U_R
=
\prod_{j=1}^R e^{-i a H(s_j)} \rev{,}
\label{eq:UR-product}
\end{equation}
\rev{with the kick area $a$ and the scaled kick times $s_j$ of Appendix~\ref{supp:theorem1_proof}.}
Since all factors commute, this becomes
\begin{equation}
U_R
=
\exp\!\left(
- i a \sum_{j=1}^R H(s_j)
\right).
\label{eq:UR-collapsed}
\end{equation}

\paragraph{Step 2: act on the ground space.}
Pick any ground state
\begin{equation}
|\psi_0\rangle \in \mathrm{Ran}\,P(0)\rev{,}
\label{eq:psi0-groundspace}
\end{equation}
\rev{where $P(s)$ is the projector onto the ground space of $H(s)$; superpositions of the four ground states are included.}
Because all $A_v$ and $B_p$ have eigenvalue $+1$ on the ground space, and this remains true for all $s$ since only the coefficients are varied, one has
\begin{equation}
H(s_j)|\psi_0\rangle
=
E_0(s_j)\,|\psi_0\rangle \rev{,}
\label{eq:Hs-on-groundstate}
\end{equation}
\rev{with the ground-state energy $E_0(s)=-[J_e(s)+J_m(s)]L^2$.}
Therefore,
\begin{equation}
U_R|\psi_0\rangle
=
e^{-i a \sum_j E_0(s_j)}\,|\psi_0\rangle ,
\label{eq:UR-on-groundstate}
\end{equation}
which is just a pure phase\rev{, the same for every state of the ground space, so that superpositions of ground states are left unchanged}.

\paragraph{Step 3: compute the defect density.}
The final defect density is
\begin{equation}
n_R
=
\frac{1}{|\Lambda|}
\langle \psi_0 |
U_R^\dagger D U_R
| \psi_0 \rangle
=
\frac{1}{|\Lambda|}
\langle \psi_0 | D | \psi_0 \rangle
=
0.
\label{eq:nR-zero}
\end{equation}

So in this topological commuting-projector setting,
\begin{equation}
n_R \equiv 0
\qquad
\text{for all } R \text{ and all } T.
\label{eq:plateau-zero}
\end{equation}

Therefore, the plateau exists, and its value is exactly zero. \rev{Since $D$ commutes with $H(s)$, the same holds for a continuous ramp of any duration and for a sudden quench, as stated in Sec.~\ref{sec:xy}.}

\section{Additional numerical results}
\label{supp:additional_numerics}
\begin{figure*}[t]
        \centering
        \includegraphics[width=1.0\linewidth]{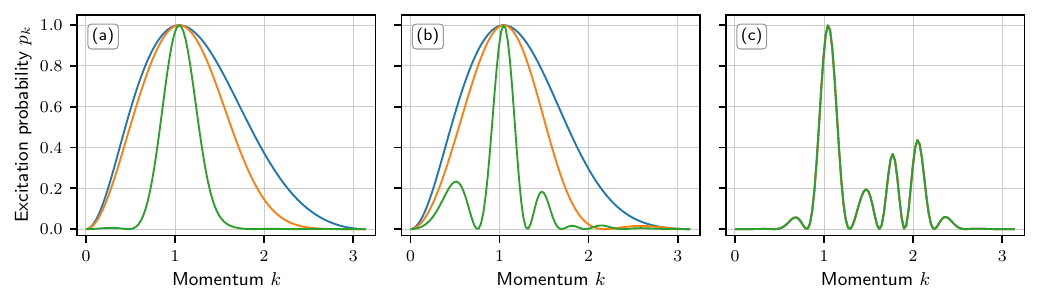}
        \caption{$p_k$ vs $k$ plots. (a) Lin strategy (b) Geo strategy (c) Geo-jump strategy for the XY model with $\gamma_i =-1, \gamma_f=1, h=0.5$ for three different total evolution \rev{times or quench rates} ($\nu=1/T$). Here, $T=[0.5, 1, 5]$ stands for the blue, orange, and green lines. In our numerics, we use a square-pulse width equal to the time vector spacing \rev{$dt$}, namely, $\Delta t = dt=0.001\,({\rm a.u.})$. The latter assures the pulses behave like single-sample kicks. In (c) the three curves coincide.
    }
        \label{fig:Pk}
\end{figure*}

Here, we show the behavior of the excitation probability at the last time of the evolution $p_k$ for different $k$ modes in the XY model, considering the lin strategy (a), the geo strategy (b), and the geo-jump strategy (c). In the geo-jump strategy, we use single-sample kicks. The geo and geo-jump strategies use the schedule shared by all modes, Eq.~\eqref{eq:closed_schedules}, with $R=5$ kicks of area $\pi/2$ at the exact scaled times $\lambda_j=(2j-1)/(2R)$, and all curves are computed on the momenta $k=(2m-1)\pi/N$ of the $N=250$ chain. The results are shown in Fig.\,\ref{fig:Pk} for final evolution times $T=[0.5,1,5]$. We see that the lin and geo strategies present similar results, while in the geo-jump strategy, all curves collapse, leading to DRIP.

\begin{figure*}[t]
        \centering
        \includegraphics[width=0.9\linewidth]{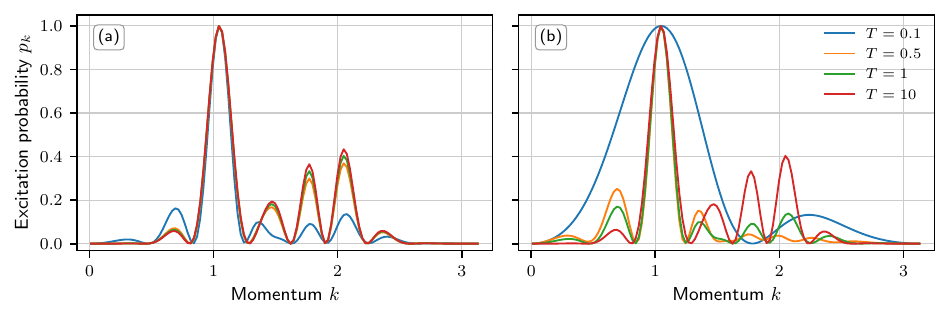}
        \caption{$p_k$ vs $k$ plots for geo-jump strategy of the XY model with $\gamma_i =-1, \gamma_f=1, h=0.5$ for four different total evolution \rev{times or quench rates} ($\nu=1/T$). Here, $T=[0.1, 0.5, 1.0, 10]$ stands for the blue, orange, green, and red lines. In our numerics, we use rectangular pulses with width $\Delta t\gg dt=0.001\,({\rm a.u.})$, namely, $\Delta t=0.01$ in (a), and $\Delta t=0.1$ in (b). Shared schedule, $R=5$; each pulse starts at $t_j=\lambda_jT$ and keeps the area $\pi/2$. In (b) the windows merge for $T=0.1$ and $T=0.5$ (blue and orange lines), which are therefore continuous drives rather than kick sequences.
    }
        \label{fig:PkDt}
\end{figure*}

\begin{figure*}[t]
        \centering
        \includegraphics[width=1.0\linewidth]{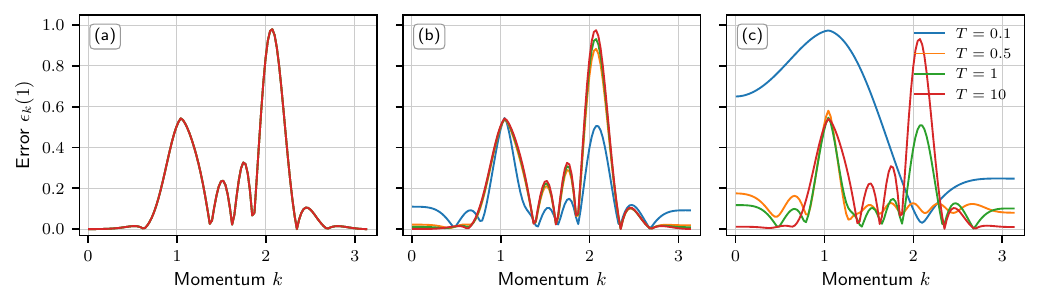}
        \caption{Error of Eq.~\eqref{eq:error} between the two eigenstates of mode $k$, $\epsilon_k(1)$, vs $k$ plots for geo-jump strategy of the XY model with $\gamma_i =-1, \gamma_f=1, h=0.5$ for four different total evolution \rev{times or quench rates} ($\nu=1/T$). Here, $T=[0.1, 0.5, 1.0, 10]$ stands for the blue, orange, green, and red lines. In our numerics, we use rectangular pulses with width $\Delta t= dt=0.001\,({\rm a.u.})$ in (a), $\Delta t=0.01$ in (b), and $\Delta t=0.1$ in (c). In (a) the four curves coincide.
    }
        \label{fig:ErrorPk}
\end{figure*}

\begin{figure}[h]
        \centering
        \includegraphics[width=1.0\linewidth]{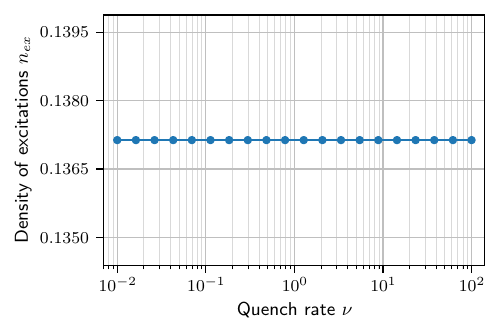}
        \caption{Rate-independent defect plateau (DRIP) for the Ising chain with $250$ spins, $h\colon10\to0$ at $\gamma=1$, with the shared schedule and $R=5$ kicks, over four decades of the quench rate. The vertical axis spans $\pm2\%$ around the plateau value $n_{ex}=0.1371$; the data are equal at every rate to the precision of the calculation. They come from the time integration of Appendix~\ref{supp:additional_numerics} with $dt=0.001$, each kick applied as a single time step of area $\pi/2$ with the control at the exact $\lambda_j$; by Appendix~\ref{supp:theorem1_proof} the result cannot depend on $T$, and the figure checks the implementation over four decades of the rate.
    }
        \label{fig:long_quench_rates_250spins}
\end{figure}

A more realistic situation involving a finite pulse width $\Delta t > dt$ will produce excitation probabilities dependent on the quench rate. This is shown in Fig.~\ref{fig:PkDt}, where we plot the excitation probabilities as a function of momentum $k$, and for different quench times $T\in [0.1,0.5,1.0,10]$ represented by the blue, orange, green, and red curves, respectively. In case (a), we use $\Delta t=0.01$, while in (b) we consider $\Delta t=0.1$. Each pulse starts at $t_j=\lambda_jT$, lasts $\Delta t$, and keeps the area $\pi/2$, while the control follows the schedule during the pulse. The rate dependence grows with $\Delta t/T$. For $\Delta t=0.1$ the five windows merge when $T\le0.5$ and the last window is cut off at $T$, so the $T=0.1$ and $T=0.5$ curves of panel (b) are continuous drives of total area $1.4$ and $7.1$ rather than kick sequences. As a complementary result, in Fig.\,\ref{fig:ErrorPk}, we show the error of Eq.~\eqref{eq:error} between the two eigenstates of mode $k$, $\epsilon_k(1)$, defined through the generalized adiabatic condition, Eq.~\eqref{eq:gac_main}.
The subfigures of \figref{fig:ErrorPk} show that the generalized adiabatic condition is not uniformly fulfilled; hence, at some $k$, the errors are quite large.
In \figref{fig:long_quench_rates_250spins}, the DRIP feature persists over a much larger range of quench rate $\nu$, from $10^{-2}$ to $10^{2}$, which confirms the rate independence summarized in Table~\ref{table:summary}: in the $\delta$-kick limit the defect density is the same at every rate to all the digits shown. Since each kick is a single time step with the control at the exact $\lambda_j$, this is a check of the implementation rather than an independent test (Appendix~\ref{supp:theorem1_proof}); rounding the kick times to the integration grid instead produces a small excursion of the plateau at $\nu\gtrsim10$.

\begin{figure*}[t]
        \centering
        \includegraphics[width=1.0\linewidth]{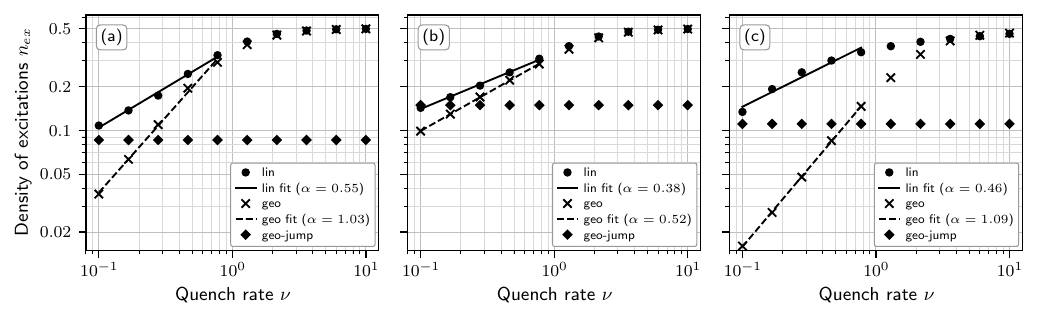}
        \caption{Same as Fig.~\ref{fig:KZ_scaling_results}, with the mode-resolved schedule of Eq.~\eqref{eq:geo_schedules} for the geo and geo-jump strategies.
    }
        \label{fig:KZ_permode}
\end{figure*}

Figure~\ref{fig:KZ_permode} repeats Fig.~\ref{fig:KZ_scaling_results} with the mode-resolved schedule of Eq.~\eqref{eq:geo_schedules}, in which every mode $k$ follows its own geodesic, $\theta_k(t)=\theta_{k,i}+(\theta_{k,f}-\theta_{k,i})t/T$. This schedule cannot be produced by a single control field of the chain, but it can be realized in the mode-by-mode emulation of Sec.~\ref{sec:outlook}, and it is the schedule for which the mode product of Appendix~\ref{supp:driven_bogoliubov} takes its simplest form. The geo-jump plateaus are again rate independent, at $0.086$, $0.149$, and $0.111$ for $R=5$, below the shared-schedule values because each mode is driven along its own shortest path. The geo data show effective exponents $\alpha=1.03$, $0.52$, and $1.09$; on the Ising line the mode-resolved geodesic gives $n_{ex}\to\nu/2\pi$ at slow rates, i.e., exponent $1$, because every mode crosses its own gap minimum at the same scaled speed.

\begin{figure*}[t]
        \centering
        \includegraphics[width=1.0\linewidth]{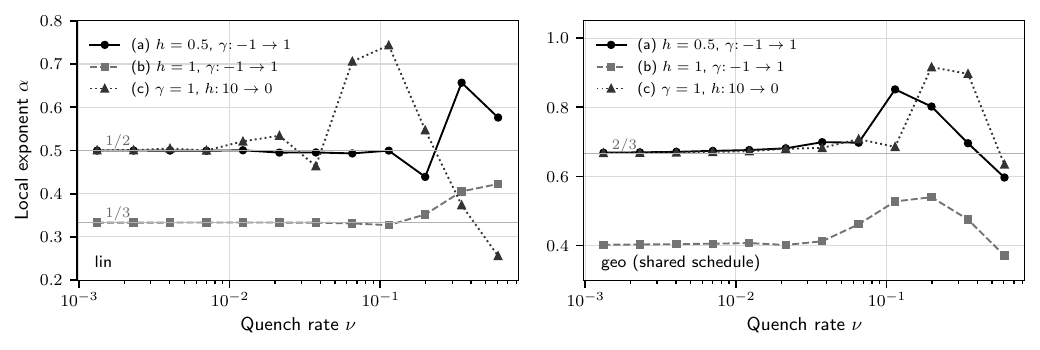}
        \caption{Local exponent $\alpha=-d\ln n_{ex}/d\ln T$ of the defect density between adjacent quench rates, for the three paths of Fig.~\ref{fig:KZ_scaling_results} in the thermodynamic limit, with the momentum grid refined near the critical mode. (a) \rev{Lin} ramps, with the asymptotic Kibble--Zurek values $1/2$ and $1/3$ as guide lines. (b) \rev{Geo} ramps with the shared schedule of Eq.~\eqref{eq:closed_schedules}, with the nonlinear-crossing value $2/3$ as a guide line; on the gapless line the exponent settles at $0.40$.}
        \label{fig:slow}
\end{figure*}

Figure~\ref{fig:slow} extends the lin and geo data of Fig.~\ref{fig:KZ_scaling_results} to slower quenches, down to $\nu=10^{-3}$, in the thermodynamic limit and with the momentum grid refined near the critical mode, and shows the local exponent $\alpha=-d\ln n_{ex}/d\ln T$ between adjacent rates. For the lin ramps it settles at $0.50$, $0.33$, and $0.50$ below $\nu\simeq5\times10^{-3}$, and for the geo ramps at $0.67$, $0.40$, and $0.67$ (Appendix~\ref{supp:KZM}). In the window $0.1\le\nu\le0.8$ of Fig.~\ref{fig:KZ_scaling_results} all six curves are still in the crossover, which is where the effective exponents quoted in Sec.~\ref{sec:xy} come from.

\clearpage
\bibliography{ref}

\end{document}